\documentclass{aa} 
\usepackage{graphicx}
\usepackage{eso-pic}    
\usepackage{color}
	\definecolor{unipd}{HTML}{b5121b}
	\definecolor{cadmiumgreen}{RGB}{2,134,121}
	\definecolor{blue}{RGB}{0,32,169}
\usepackage{multirow}
\usepackage{pifont}

\usepackage{abraces}
\usepackage{relsize}
\usepackage{txfonts}
\usepackage{mathtools}
\usepackage{hyperref}
   \hypersetup{colorlinks=true, linkbordercolor=blue, breaklinks=true, citecolor=blue, filecolor=black, linkcolor=blue, urlcolor=blue}
\begin{document}

   \title{Forming merging double compact objects with stable mass transfer}


   \author{A. Picco
          \inst{1}
          \and
          P. Marchant\inst{1}
          \and
          H. Sana\inst{1}
         \and 
         G. Nelemans\inst{2}
          }

   \institute{Institute of Astronomy, KU Leuven, Celestijnlaan 200D, 3001 Leuven,
        Belgium\\
              \email{annachiara.picco@kuleuven.be}
         \and
         Dept. Astrophysics / IMAPP,
         Radboud University Nijmegen, Nijmegen, NL
             }

   \date{Last compiled: \today}

 
  \abstract
   {Merging double compact objects (CO) are one of the possible endpoints of the evolution of stellar binary systems. As they represent the inferred sources of \text{every} detected gravitational wave (GW) signal, modeling their progenitors is of paramount importance both to gain a better understanding of gravitational physics and to constrain stellar evolution theory.}
   {Stable mass transfer (MT) between a donor star and a black hole is one of the proposed tightening mechanisms to form binary black holes that merge within the lifetime of the universe. We aim to assess the potential of stable non-conservative mass transfer to produce different pairings of compact objects including black holes (BHs), neutron stars (NSs) and white dwarfs (WDs).}
   {We investigate the conditions (orbital periods and mass ratios) required for mass transfer between a star and a CO to be stable and to lead to binary COs that merge within a Hubble time. We use published results for the response of the stellar radii to rapid mass loss, covering different evolutionary stages and masses. Coupled with analytical models for orbital evolution, we determine the boundary for unstable mass transfer as well as the post-interaction properties of binaries undergoing stable MT. In addition, we investigate the impact of the angular momentum loss prescription in the resulting hardening by accounting for both the isotropic re-emission from the accretor’s vicinity and mass outflow from the second Lagrangian point.}
   {Stable MT in systems with a CO + Roche lobe-filling star, in the completely non-conservative limit of isotropic re-emission from the vicinity of the accretor, is shown to be able to form any pair of merging double COs, with the exception of WD + BH and with a limited parameter space for NS + NS. Considering the possibility of mass outflow from the Lagrangian point $L_2$, the resulting parameter space for GW progenitors is shifted towards smaller initial mass ratios (defined as the ratio of the donor mass over the CO mass), consequently ruling out the formation of NS + NS pairs while allowing the production of merging WD + BH binaries. We compare our results with observations of single-degenerate binaries and find that the conditions for the stable MT channel to operate are present in nature. We then show that stable mass transfer in the isotropic re-emission channel can produce merging binary BHs with mass ratios $>0.1$, consistent with the majority of inferred sources of the third Gravitational Wave Transient Catalogue. Enhanced angular momentum loss from $L_2$ lifts the minimum final mass ratio achievable by stable mass transfer.}
   {}


   \maketitle
%

\section{Introduction}

Ground-based gravitational wave (GW) observatories of the LIGO, Virgo and KAGRA (LKV) network have successfully identified coalescing compact objects originating from binary black holes (BHs), binary neutron stars (NSs) and binaries containing both a BH and a NS. Signals from these sources are compiled in the third gravitational wave transient catalogue (GWTC-3, \citealt{Abbott2021III}) and currently form a sample of over 90 detections, expected to grow significantly with future observing runs. In the next decade, third generation detectors such as the Einstein Telescope (\citealt{Maggiore2020}) will provide hundreds of thousands of detections per year (\citealt{Hall2019}), and the Laser Interferometer Space Antenna (\citealt{LISA2017}) will open up the frequency range of merging binary white dwarfs (WD).

While the observed GW sample is increasing and progressively revealing new types of sources, our theoretical understanding of the evolutionary channels that form them is still limited. The main formation pathway leading to BH + BH and BH + NS mergers is still a matter of debate. On one hand, dynamical interactions in globular clusters (\citealt{Kulkarni1993}, \citealt{Sigurdsson1993}, \citealt{Portegies2000}), young stellar clusters (e.g. \citealt{Mapelli2016}) or in isolated hierarchical systems (e.g. \citealt{Thompson2011}) are predicted to contribute in forming close double BHs. On the other hand, isolated binary evolution scenarios rely either on a process that could prevent stars in very close binaries from expanding, like chemically homogeneous evolution (\citealt{Marchant2016}, \citealt{Mandel2016}), or on a hardening mechanism allowing wide binaries to come close enough. 

Two main hardening mechanisms have been proposed for isolated binaries: evolution of the binary through a common envelope (CE) phase (\citealt{Paczynski1976}) or a stable mass transfer (MT) episode via Roche lobe overflow (\citealt{vdHeuvel2017}). The CE scenario was first contemplated in the context of the evolution of single-degenerate binaries with WD objects (e.g. \citealt{Sparks1974}, \citealt{Chau1974}), then introduced to explain the origin of cataclysmic variable stars (\citealt{Webbink1975}), and subsequently suggested as the origin of Type I supernovae (\citealt{Webbink1984}). CE was also advanced as the standard, late-evolutionary channel towards double degenerate binaries, both in the context of the final evolution of massive  X-ray binaries as well as in the formation of NS binaries (\citealt{vdHeuvel1976}) and double WD (e.g. \citealt{Iben1984}). During such CE phase, the components of the binary system are engulfed by a common gaseous envelope; the resulting drag dissipates energy (and momentum) and dramatically shrinks the orbit. This process is expected to happen on a dynamical timescale, and the survival of the shrinked binary is subject to the envelope structure and companion masses. 

Conversely, \textit{stable mass transfer} via Roche lobe overflow implies a quiet, long lived phase of tightening (or broadening) of the system’s orbit. This happens on the thermal or nuclear timescale of the donor star and depends on the evolutionary stage of the donor itself. During a stable mass transfer episode, the donor keeps its radius close to that of its Roche lobe, while material from its envelope escapes the star's gravitational well and is transferred to the companion. The MT rate is governed by the response of the donor's own radius to mass loss, and by the Roche lobe radius response. The mechanism is self-regulating (i.e. a maximum mass transfer rate is expected to be reached and, subsequently, quenched), thus dubbed stable. 

While stable MT is usually involved, in interacting non-degenerate binary systems, before the formation of the first CO (e.g., \citealt{Dominik2012}, \citealt{Langer2020}), CE evolution has been considered the classical hardening channel towards the merging double-degenerate configuration, after the first CO is already formed (e.g. \citealt{Belczynski2016}, \citealt{Liu2022}). However, \cite{vdHeuvel2017} recently argued that the role of stable MT might be crucial in tightening the orbit of systems composed by a donor and a CO. This evolutionary scenario hypothesizes the MT episode to be completely non-conservative, and the avoidance of a CE episode to be dependent on the mass ratio of the system. 

More specifically, the complete non-conservativeness assumption has been translated in the past into modeling mass leakage from the system from the vicinity of the first born CO with its specific angular momentum, in the \textit{isotropic re-emission} channel (see \citealt{vdHeuvel2017}), usually justified by super-Eddington accretion rates in proximity of a compact accretor. However, the isotropic re-emission channel is just one possibility. At very high mass transfer rates, systems can potentially experience outflows of mass through the $L_2$ Lagrangian point \citep{Lu2023}, which would lead to further orbital shrinkage: the presence of a circumbinary outflow from systems such as the galactic microquasar SS433 could support this possibility \citep{Fabrika1993}.

Population synthesis techniques are key to understand the increasing sample of merging double COs via any kind of evolutionary channel, and recent studies have shown that the contribution from stable, non-conservative MT to the observed sample of merging binary BHs can be substantial and even larger than that from the CE channel (\citealt{Neijssel2019}, \citealt{Bavera2021}). There are however uncertainties in the critical mass ratios for CE. As the response of a star to mass loss depends on the structure of its envelope, several classical (e.g. \citealt{Paczynski1965}) and more modern (e.g. \citealt{Pavlovskii2015}) studies made use of stellar structure models to survey ranges of mass ratios and orbital periods for (un)stable MT; their results are subsequently implemented in binary population synthesis codes as physically motivated criteria to identify CE. In particular, population synthesis studies usually assume a direct correlation between a specific stellar evolutionary stage with the (un-)successful CE survival, see e.g. the mass ratio thresholds from \cite{Hjellming1987} and \cite{Hurley2002}. This approach gives an overestimate of the CE channel contribution to binary BHs \citep{Marchant2021}. Furthermore, the adopted critical mass ratios for CE evolution may lead to an over-simplification of the progenitors' initial conditions to produce merging double COs \citep{GallegosGarcia2021}.

The aim of this work is to assess the potential of non-conservative stable MT as a candidate evolutionary channel towards the merging double CO configuration. We will put the emphasis on the production of \textit{all the different pairings} of compact objects other than just the double BH scenario. We do so by investigating the parameter space of progenitor systems composed by a CO orbiting a Roche lobe-filling donor, using existing results for the response to a star to rapid mass transfer \citealt[\citeyear{Ge2015}, \citeyear{Ge2020}]{Ge2010} to resolve the boundary of critical mass ratios for CE evolution. We then compute the orbital shrinkage of the systems with angular momentum balance arguments and finally study the possibility for the formation of the different GW sources pairings. Crucially, our approach is semi-analytical, as it does not rely on the tools of population synthesis and/or detailed binary interaction simulations. Section \ref{sec:Methodology} provides the methodology; we present our results for selected donor star's masses in Sect. \ref{sec:Results}, while Sect. \ref{sec:4_Comparison with observations} compares the results with observations. We conclude by summarizing the main findings and discussing remarks / perspective work in Sect. \ref{sec:Conclusions}.

\section{Methodology}\label{sec:Methodology}
We will always refer to quantities concerning the donor (accretor) in the MT episode with the subscript d (a), and with i (f) to quantities at the start (end) of the MT episode. If neither i nor f is specified, we refer to quantities as variables along the evolution of the systems. We define the binary mass ratio $q$ between a donor of mass $m_{\mathrm{d}}$ and an accretor of mass $m_{\mathrm{a}}$, at an arbitrary stage, as 
\begin{equation}
    q\equiv\dfrac{m_\mathrm{d}}{m_\mathrm{a}}\:.
\end{equation}
We also define a stripping factor, $\alpha_{\mathrm{core}}$, as the fraction of initial donor's mass, $m_{\mathrm{d,i}}$, which remains after the MT episode, leaving the donor with a mass $m_{\mathrm{d,f}}$:
\begin{equation}
\alpha_{\mathrm{core}}\equiv\dfrac{m_{\mathrm{d,f}}}{m_{\mathrm{d,i}}}<1\:.
\end{equation}
Therefore, the final mass ratio of the system will be
\begin{equation}\label{eq:final_mass_ratio}
q_{\mathrm{f}}=\dfrac{m_{\mathrm{d,f}}}{m_\mathrm{a,f}}=\dfrac{\alpha_{\mathrm{core}}\:m_{\mathrm{d,i}}}{m_\mathrm{a,f}}\:.
\end{equation}
Section \ref{sec:2_1_shrinkage} and Sect. \ref{sec:2_2_mergingtime} set the hypothesis on the shrinking mechanism and the merging time calculation, presenting also a functional study for a simple case of fixed donor mass $m_{\mathrm{d,i}}=32\:M_{\odot}$; Sect. \ref{sec:2_3_commonenvelope} discusses the instability boundary calculation, with a description on the properties used from existing simulations in Sect. \ref{sec:2_4_simulations}; Sect. \ref{sec:Final_orbital_separation} sets the constraint on the final donor star's radius; Sect. \ref{sec:L2_outflow}  shows how we model mass outflow from $L_2$.

\subsection{The parameter space for GW progenitors}\label{sec:2_1_shrinkage}
We are hypothesizing a completely non-conservative MT episode: the first-born CO is assumed to be accreting material from the donor at a super-Eddington rate, thus expelling mass from its surroundings with the accretor's specific angular momentum. This scenario has been referred to as \emph{isotropic re-emission}.

We define the efficiency $\epsilon$ of the MT, i.e. the fraction of accreted mass $\Delta m_{\mathrm{a}}$ when the donor transfers $\Delta m_{\mathrm{d}}$, as
\begin{equation}\label{eq:epsilon_uguale_uno_meno_beta}
\epsilon=\dfrac{\Delta m_{\mathrm{a}}}{\Delta m_{\mathrm{d}}}\equiv 1-\beta\:.
\end{equation}
Here $\beta=0$ is the limit of conservativeness (everything is accreted) and $\beta=1$ describes the isotropic re-emission assumption. Assuming that $\beta$ is a constant of evolution, an exact orbital solution for the period shrinkage $P_{\mathrm{i}}/P_{\mathrm{f}}$ can be found by angular momentum balance arguments \citep{Soberman1997}:
\begin{equation}\label{eq:ratio_isotropic_reemission}
   \dfrac{P_{\mathrm{i}}}{P_{\mathrm{f}}}=e^{-3\left(q_{\mathrm{f}}-q_{\mathrm{i}}\right)}\left(\dfrac{q_{\mathrm{f}}}{q_{\mathrm{i}}}\right)^3\left(\dfrac{1+q_{\mathrm{f}}}{1+q_{\mathrm{i}}}\right)^2\:,
\end{equation}
where, within our assumption of $\beta=1$, the accretor mass will satisfy $m_{\mathrm{a,i}}=m_{\mathrm{a,f}}$, so that the final and initial mass ratio of the system are related by $q_{\mathrm{f}}=\alpha_{\mathrm{core}}\:q_{\mathrm{i}}$.

To assess if the binary system at the end of the MT can evolve into a GW source, we adopt the analytical expression for the delay time $t_{\mathrm{merge}}$ of an inspiraling binary (\citealt{Peters1964}),
\begin{equation}\label{eq:Peters}
   t_{\mathrm{merge}}=\dfrac{a_{\mathrm{f}}^4}{4\mathcal{B}}\hspace{0.5cm}\mathrm{with}
\end{equation}
\begin{equation}
\mathcal{B}\equiv\dfrac{64}{5}\dfrac{G^3}{c^5}(m_{\mathrm{a,f}}+m_{\mathrm{d,f}})\:m_{\mathrm{a,f}}\: m_{\mathrm{d,f}}\:,
\end{equation}
which holds for a circular orbit with final orbital separation $a_{\mathrm{f}}$ and neglects other sources of radiation other than the gravitational quadrupole one. We are, therefore, implicitly assuming that the second born CO, at the start of the inspiral phase, has the post-MT mass $m_{\mathrm{d,f}}$ and starts the inspiraling at the post-MT separation $a_{\mathrm{f}}$. We shall relax this assumption in Sect. \ref{sec:NSdonor}. 

\subsection{Constant $\alpha_{\mathrm{core}}$}\label{sec:2_2_mergingtime}
The initial conditions required for a GW progenitor system can be constrained by looking at the initial accretor mass vs initial orbital period,  $(m_{\mathrm{a,i}}-\log P_{\mathrm{i}})$, parameter space. To do so, we combine Eq. \ref{eq:ratio_isotropic_reemission} with Eq. \ref{eq:Peters} and Kepler's third law for the final orbital separation $a_{\mathrm{f}}$. By imposing the merging time being less than a Hubble time $t_H$, i.e. $t_{\mathrm{merge}}<t_{H}\simeq 13.8$ Gyr, one finds
\begin{equation}\label{eq:isotropic_reemission_condition}
   P_{\mathrm{i}}(m_{\mathrm{d,i}},m_{\mathrm{a,i}},t_{\mathrm{merge}},\alpha_{\mathrm{core}})<
\end{equation}
\[0.642\:\left(\dfrac{t_{\mathrm{merge}}}{t_{H}}\right)^{3/8}\left(\dfrac{m_{\mathrm{a,i}}\:\alpha_{\mathrm{core}}\:m_{\mathrm{d,i}}}{M_{\odot}^2}\right)^{3/8}\left(\dfrac{m_{\mathrm{a,i}}+\alpha_{\mathrm{core}} m_{\mathrm{d,i}}}{M_{\odot}}\right)^{-1/8}\]
\[\times \:\underbrace{e^{-3\left(q_{\mathrm{f}}-q_{\mathrm{i}}\right)}\left(\dfrac{q_{\mathrm{f}}}{q_{\mathrm{i}}}\right)^3\left(\dfrac{1+q_{\mathrm{f}}}{1+q_{\mathrm{i}}}\right)^2}_{=P_{\mathrm{i}}/P_{\mathrm{f}}}\:\mathrm{days}\:,\]

This condition can be generalized to other hardening channels if the value of $P_{\mathrm{i}}/P_{\mathrm{f}}$ is a function of the initial and final mass ratios. We will use this later when we consider outflows from the $L_2$ point.

\begin{figure}
   \includegraphics[width=0.475\textwidth]{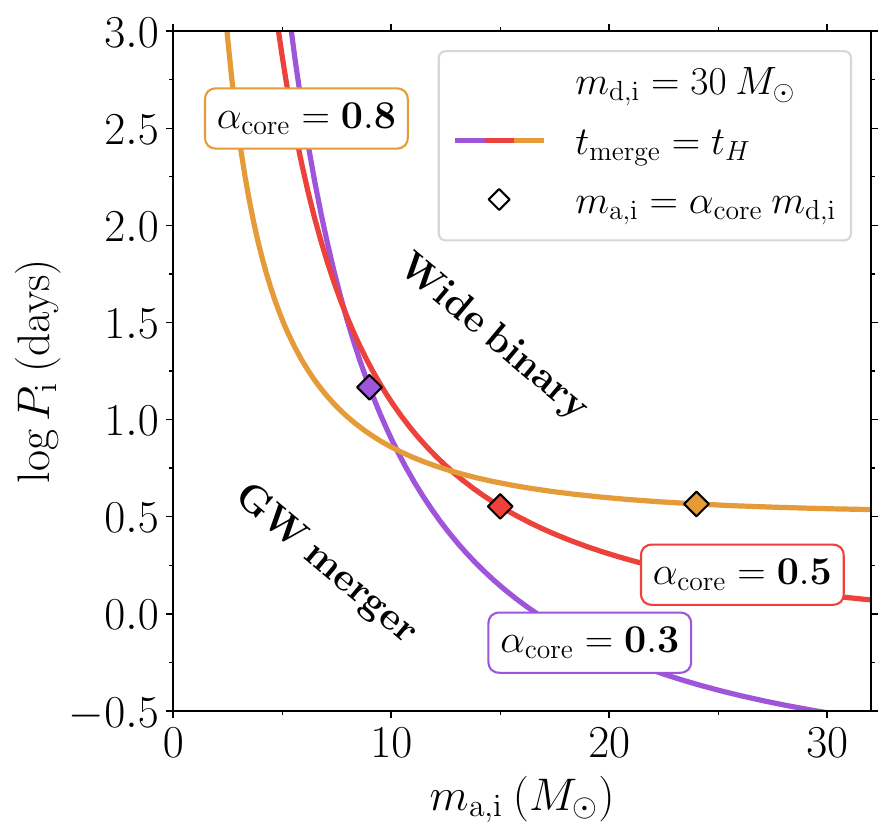}
      \caption[]{Locus (see Eq. \ref{eq:isotropic_reemission_condition}) for the formation of a GW source through stable mass transfer with a $30\:M_\odot$ donor as a function of the CO accretor mass $m_\mathrm{a,i}$ and orbital period $P_\mathrm{i}$. The solid curves represent the upper bounds on initial periods $\log P_{\mathrm{i}}$ to achieve a delay time of $t_{\mathrm{merge}}=t_{H}$; different stripping factors $\alpha_{\mathrm{core}}$ are labeled along the corresponding curves. Colored diamonds correspond to the point, along their respective curves, in which the MT episode is expected to invert the mass ratio.}
      {\label{fig:logPP0_first}}
   \end{figure}

A visualization of the locus for fixed values of the stripping factor $\alpha_{\mathrm{core}}$ and a donor initial mass of $m_{\mathrm{d,i}}=30 \:M_{\odot}$ is shown in Fig. \ref{fig:logPP0_first}. The parameter space which stands below the solid curves is the promising portion for the initial properties of progenitor systems of GW sources. Systems with larger initial orbital period $P_{\mathrm{i}}$ are more likely to originate sufficient shrinkage for a GW merger at large initial mass ratios, $q_{\mathrm{i}}\gg 1$. These systems are also farther from reaching the mass ratio inversion (around $q=1$, the colored diamonds along the curves) during the MT episode, thus less likely to switch from orbital contraction to orbital expansion due to MT. The effect of different stripping factors $\alpha_{\mathrm{core}}$, labeled along the correspondent curves, is also shown: we see that more stripping (moving from the $\alpha_{\mathrm{core}}=0.5$ to the $\alpha_{\mathrm{core}}=0.3$ curve) implies less favorable parameter space for GW mergers after the mass inversion $q=1$, as the broadening of the orbit during the MT is favored. On the other hand, less stripping of the donor (moving from the $\alpha_{\mathrm{core}}=0.5$ to the $\alpha_{\mathrm{core}}=0.8$ curve) allows for less extreme $q_{\mathrm{i}}$ to form GW mergers.

\subsection{The common envelope boundary}\label{sec:2_3_commonenvelope}
The assumption of a constant stripping factor is useful for illustrative purposes, but we want to trace a more accurate boundary in the parameter space $(m_{\mathrm{a,i}}-P_{\mathrm{i}})$ of GW progenitors. To this purpose, we also want to take into account that, keeping a fixed donor mass $m_{\mathrm{d,i}}$, for more extreme initial mass ratios we expect to encounter the boundary of critical mass ratios $q_{\mathrm{crit}}$ for CE evolution \citep{Soberman1997}. 

The (in-)stability of a MT episode is usually characterized with mass-radius exponents. We follow \cite{Soberman1997} to write the donor's mass-radius critical (or adiabatic) exponent $\zeta_{\mathrm{ad}}$:
\begin{equation}\label{eq:zeta_ad}
   \left.\zeta_{\mathrm{ad}}\equiv\dfrac{\partial\log R_{\mathrm{d}}}{\partial\log m_{\mathrm{d}}}\right|_{\mathrm{ad\: MT}}\:.
 \end{equation}
From the calculations of \cite{Ge2020} we take $\zeta_\mathrm{ad}(R,M)$, that is the adiabatic mass radius exponent as a function of the stellar radius (encoding its evolutionary stage) and mass (see more in Sect. \ref{sec:2_4_simulations}).
The orbit response to the MT episode is quantified by the donor's Roche lobe mass-radius exponent, $\zeta_{RL}$:
\begin{equation}
\zeta_{RL}\equiv\dfrac{\partial \log R_{RL}}{\partial\log m_{\mathrm{d}}}\:.
\end{equation}
Performing the stability analysis for the adiabatic MT episode amounts to compare the two dimensionless exponents $\zeta_{RL}$ and $\zeta_{\mathrm{ad}}$. Stability requires that, after a mass loss-induced perturbation, $\delta m_{\mathrm{d}}<0$, the star's radius remains contained within its Roche lobe; i.e. $\zeta_{\mathrm{ad}}>\zeta_{RL}$. If this is not satisfied, the MT is unstable and we expect CE evolution to happen. 

The donor's Roche lobe mass-radius exponent $\zeta_{RL}$ must be computed according to the evolution of the binary system, so an assumption on the MT mode is needed; indeed, analytical expressions are provided in \cite{Soberman1997}. In the case of conservative MT, one derives 
\begin{equation}\label{eq:zeta_cons}
   \zeta_{RL}^{\mathrm{cons}}=-2+2q+\dfrac{\partial\log(R_{RL}/a)}{\partial\log q}\:,
\end{equation}
where we adopt the \cite{Eggleton1983} prescription for $R_{RL}$:
\begin{equation}\label{eq:Eggleton}
   \dfrac{R_{RL}}{a}=\dfrac{0.49 q^{2/3}}{0.6 q^{2/3}+\log(1+q^{1/3})}\:,\hspace{0.5cm}0<q<\infty\:,
\end{equation}
with $a$ being the orbital separation. 

If we include non-conservativeness, $\epsilon<1$, by switching on an efficiency $\beta>0$ for isotropic re-emission as in Eq. \ref{eq:epsilon_uguale_uno_meno_beta}, the Roche lobe mass-radius exponent $\zeta_{RL}^{\epsilon=1-\beta}$ assumes the following general form:
\begin{equation}\label{eq:zeta_uno_meno_beta}
   \zeta_{RL}^{\epsilon=1-\beta}=
\end{equation}
\[-2+\dfrac{2q\:(1+\epsilon q)}{q+1}-(1-\epsilon)\dfrac{q}{1+q}+\beta\dfrac{2 q^2}{q+1}+\dfrac{\partial\log(R_{RL}/a)}{\partial\log q}\:.\]
In the assumption of fully non-conservative MT, i.e. $\epsilon=0$, and purely isotropic re-emission, i.e. $\beta=1$, the above expression translates into
\begin{equation}\label{eq:zeta_iso}
\zeta_{RL}^{\mathrm{iso}}=\zeta_{RL}^{\mathrm{cons}}-\dfrac{q}{q+1}\:,
\end{equation}
where the superscript `$\mathrm{iso}$` stands for isotropic re-emission. 

By comparing the adiabatic mass-radius exponent of donor stars, $\zeta_\mathrm{ad}(R_\mathrm{d,i},m_\mathrm{d,i})$, with a family of Roche lobe mass-radius exponents, $\zeta_{RL}^{\mathrm{iso}}=\zeta_{RL}^{\mathrm{iso}}(q)$, sharing the same initial mass $m_{\mathrm{d,i}}$ and radius $R_{\mathrm{d},\mathrm{i}}$, the threshold in mass ratio for which MT is unstable, $q_{\mathrm{crit}}^{\mathrm{iso}}$, is found by requesting
\begin{equation}\label{eq:stability_isotropic_re-emission_condition}
\zeta_{\mathrm{ad}}(R_\mathrm{d,i},m_\mathrm{d,i})=\zeta_{RL}^{\mathrm{iso}}(q_{\mathrm{crit}}^{\mathrm{iso}})\:,
\end{equation}
where we called $q_{\mathrm{crit}}^{\mathrm{iso}}$ the critical mass ratio of an interacting binary system at the start of the unstable MT episode. 

Once we derived critical mass ratios, we can build a common envelope boundary in the $(m_{\mathrm{a,i}}- P_{\mathrm{i}})$ parameter space, for fixed $m_{\mathrm{d,i}}$, with the following reasoning. Let us consider a donor star of radius $R_{\mathrm{d,i}}$, which we assume equal to the Roche lobe radius $R_{RL,\mathrm{i}}$ at the onset of MT. The value of $\zeta_{\mathrm{ad}}$ of the star at the evolutionary stage in which it fills its Roche lobe yields then the critical mass ratio $q_{\mathrm{crit}}^{\mathrm{iso}}$ for unstable MT, by solving Eq. \ref{eq:stability_isotropic_re-emission_condition} (see also Fig. \ref{fig:zetas}). Combining the information of $q_{\mathrm{crit}}^{\mathrm{iso}}$ and $R_{\mathrm{d,i}}=R_{RL,\mathrm{i}}$, we can then infer a critical period for the instability boundary using Kepler's third law:
\begin{equation}\label{eq:common_envelope_boundary}
P_{\mathrm{i, crit}}=\dfrac{2\pi}{\left[G(m_{\mathrm{d,i}}+m_{\mathrm{a,i}})\right]^{1/2}}\left.\left(\dfrac{R_{RL}}{a}\right)\right|_{q^{\mathrm{iso}}_{\mathrm{crit}}}^{-3/2}R_{\mathrm{d,i}}^{3/2}\:.
\end{equation}
Here we isolated the factor $R_{RL}/a$, which is immediately known at the wanted mass ratio $q=q_{\mathrm{crit}}^{\mathrm{iso}}$ (see Eq. \ref{eq:Eggleton}). Notice that this factor is evaluated at $q_{\mathrm{crit}}^{\mathrm{iso}}$, i.e. at the start of the MT episode. We determine $\zeta_{\mathrm{ad}}$ as a function of donor stars' initial mass $m_{\mathrm{d,i}}$ and radius $R_{\mathrm{d,i}}$ we use the simulations of \cite{Ge2020}, described in more detail in Sect. \ref{sec:2_4_simulations}.

\subsection{Ge et al. simulations}\label{sec:Ge_simulations}\label{sec:2_4_simulations}
\begin{figure}
   \centering
   \includegraphics[width=0.475\textwidth]{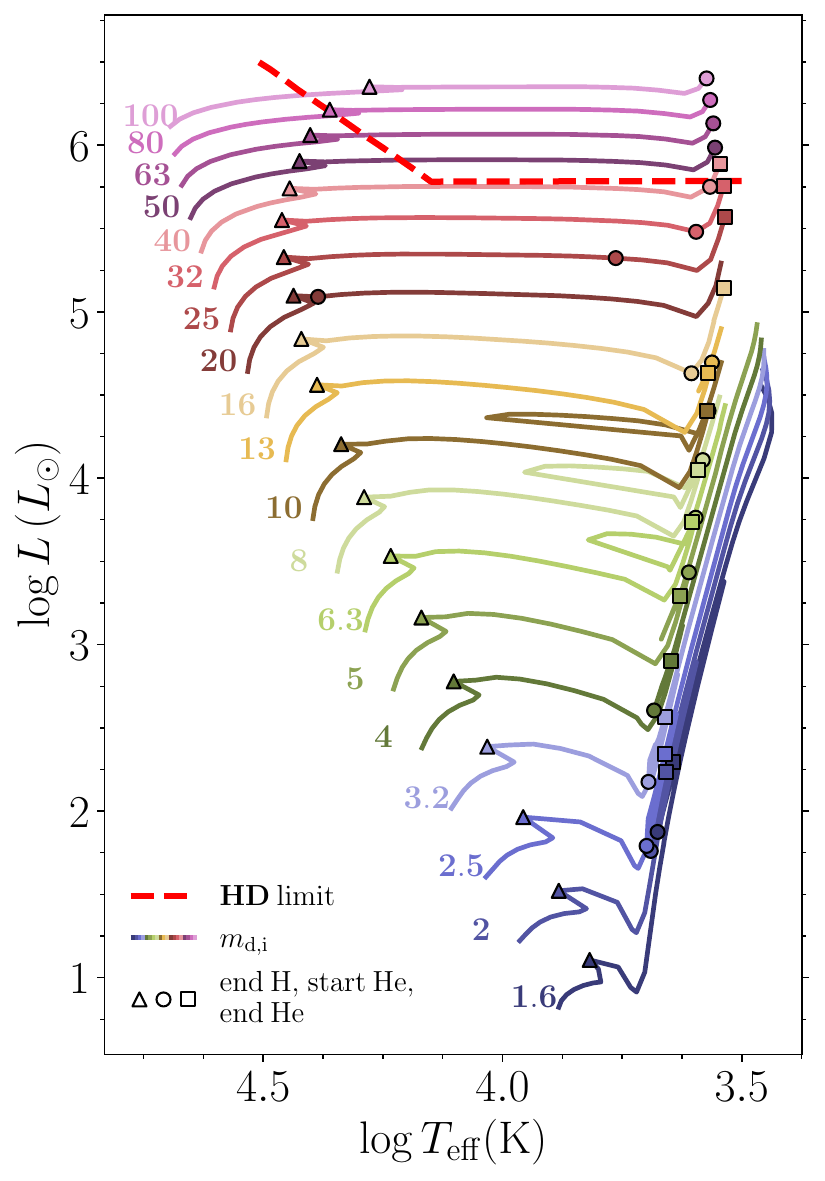}
      \caption[]{Hertzsprung-Russel diagram of the adiabatic mass loss sequences from \cite{Ge2020}, labeled by their mass. Different evolutionary phases (TAMS, Helium ignition and end of core He-burning) are highlighted with different icons (unfilled triangle, circle and square). The empirical Humphrey-Davidson (HD) limit, in dashed red \citep{HumphreysDavidson1979} is also reported.}
      {\label{fig:HR}}
   \end{figure}

The set of simulations from \cite{Ge2020} covers a grid of Population I stars, $Z=0.02$, spanning the Zero Age Main Sequence (ZAMS) mass range $0.10\leq m_{\mathrm{d,i}}/M_{\odot}\leq 100$ and evolutionary stages from ZAMS through the Terminal Age Main Sequence (TAMS), to the Helium ignition, Red Giant Branch (RGB) and Asymptotic Giant Branch (AGB) as the case may be. Following the methodology described in \cite{Ge2010}, they build \textit{adiabatic mass loss sequences}: starting from an initial stellar model at an arbitrary state in its evolution, mass is removed from the surface while the stellar interior relaxes adiabatically to a new state of hydrostatic equilibrium. These adiabatic mass loss sequences yield $\zeta_{\mathrm{ad}}$, fixing the critical mass ratio for dynamical timescale mass transfer. 
The Hertzsprung-Russel diagram of the mass loss sequences, for $1.6\leq m_{\mathrm{d,i}}/M_{\odot}\leq 100$, is shown in Fig. \ref{fig:HR}.

We use the critical mass-radius exponents $\zeta_{\mathrm{ad}}$ from \cite{Ge2020} to derive the threshold for unstable MT using Eq. \ref{eq:stability_isotropic_re-emission_condition}. Their mass-radius relations' slopes are derived in the fully conservative approximation, namely by assuming a Roche lobe response as in Eq. \ref{eq:zeta_cons}. They nevertheless claim that their exponents can be used also in the non-conservative cases to find critical mass-ratios for CE (see \citealt{Ge2010}). 

From their simulations we also take the core mass, defined as the mass coordinate at which the helium abundance is halfway between the surface helium abundance and the maximum helium abundance in the stellar interior. We assume this $m_{\mathrm{c}}$ to be the remaining mass after the MT episode, so that our stripping factor $\alpha_{\mathrm{core}}$ becomes
\begin{equation}
\alpha_{\mathrm{core}}\equiv\dfrac{m_{\mathrm{d,f}}}{m_{\mathrm{d,i}}}=\dfrac{m_{\mathrm{c}}}{m_{\mathrm{d,i}}}\:.
\end{equation}
Thus, if the binary system survives the MT episode, we assume it to be composed by an accretor with mass $m_{\mathrm{a,f}}$ and a donor with $m_{\mathrm{d,f}}=m_{\mathrm{c}}$. There is a subtlety though: during the MS, $m_{\mathrm{c}}$ as defined by \cite{Ge2020} recedes over a nuclear timescale. This receding resembles the monotonous one, predicted by single star evolution, of the convective core of massive stars. Adopting such $m_{\mathrm{c}}$ before TAMS can thus lead to a consistent over-estimation of the core left after the MT episode, as a Case A MT is expected to quickly strip the donor and make the core retract deeper. In this respect, we model the remaining mass as follows:
\[
   \begin{matrix*}[l]
          \text{\textsc{Pre-TAMS}}:& \hspace{0.75cm}m_{\mathrm{d,f}}=\left. m_{\mathrm{c}}\right|_{\mathrm{TAMS}}, \\
          \text{\textsc{Post-TAMS}}:& \hspace{0.75cm}m_{\mathrm{d,f}}=m_{\mathrm{c}}\:,
   \end{matrix*}\hspace{0.5cm}
\]
namely we are using the core mass at TAMS, $\left. m_{\mathrm{c}}\right|_{\mathrm{TAMS}}$, as the remaining mass $m_{\mathrm{d,f}}$, after MT episodes initiated by pre-TAMS donors. This is still expected to be an overestimate on the final mass after mass transfer.

For the largest initial donor star masses, namely $50\leq m_{\mathrm{d,i}}/M_{\odot}\leq 100$, we limit our analysis to those initial models in the mass loss sequences that fall under the empirical Humphrey-Davidson limit (dashed red in Fig. \ref{fig:HR}).

\begin{figure}
   \includegraphics[width=0.475\textwidth]{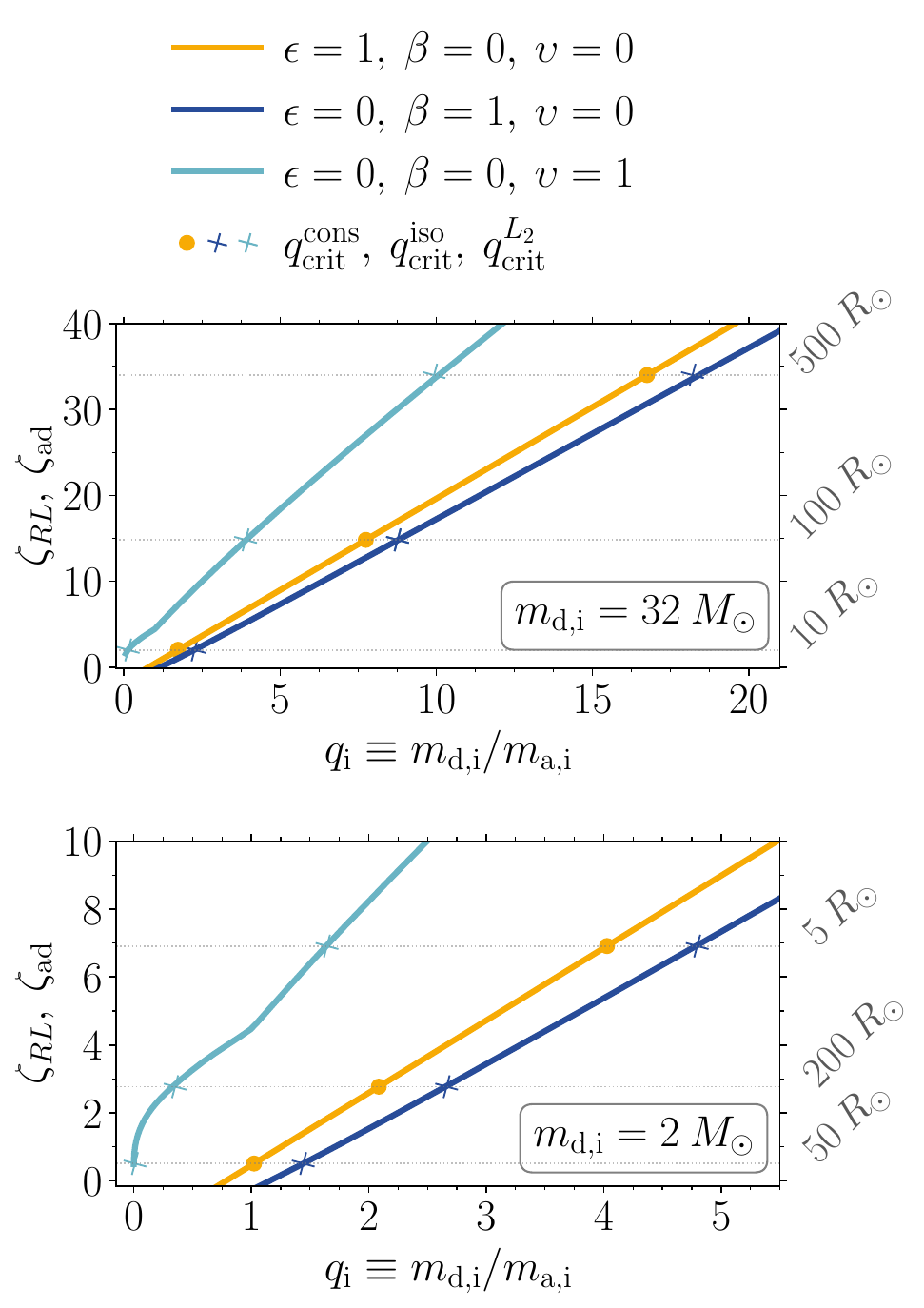}
      \caption[]{Functional dependence of $\zeta_{RL}$ on the binary mass ratio $q_{\mathrm{i}}\equiv m_{\mathrm{d,i}}/m_{\mathrm{a,i}}$, compared to the adiabatic index $\zeta_{\mathrm{ad}}$ for donor stars of $32\:M_\odot$ (top panel) and $2\:M_\odot$ (bottom panel) of different radii. The different colors of the solid lines show the conservative limit $\epsilon=1$ (golden) and the completely non conservative cases of purely isotropic re-emission $\beta=1$ (dark blue) and mass outflow from $L_2$ (dark light blue), see the discussion in Sect. \ref{sec:L2_outflow}. Critical mass ratios $\zeta_{\mathrm{ad}}$ for the MT to be unstable as calculated by \cite{Ge2020} are shown as horizontal lines (labeled by the donor star radius). For each radius and mode of mass transfer, the critical mass ratio corresponds to the crossing points between $\zeta_{\mathrm{ad}}$ and $\zeta_{RL}$.}
      {\label{fig:zetas}}
   \end{figure}

Figure \ref{fig:zetas} shows how the critical mass ratios for unstable mass transfer are determined using two fixed donor masses from the \cite{Ge2020} grid. At any evolutionary stage, the critical mass ratios for isotropic re-emission $q^{\mathrm{iso}}_{\mathrm{crit}}$ are shifted, with respect to the conservative case, towards smaller values (i.e., to larger accretor masses $m_{\mathrm{a,i}}$). Massive stars experience significant radial expansion in the Hertzsprung gap before developing a convective envelope, while lower mass stars develop a convective envelope and start rising the Hayashi line without significant expansion after the main sequence. In terms of $\zeta_\mathrm{ad}$, this results in the more massive star experiencing a monotonic increase of $\zeta_\mathrm{ad}$ over a large range of radii, while the opposite is the case for lower mass stars.

\subsection{Final orbital separations}\label{sec:Final_orbital_separation}
We compute the final orbital separations $a_{\mathrm{f}}$ by using the orbital shrinkage equation, Eq. \ref{eq:ratio_isotropic_reemission}, and Kepler's third law.
Even if the instability criterion is satisfied, we find cases where the predicted orbital separations $a_{\mathrm{f}}$ at the end of the MT episode are too small to contain a stripped helium star of the expected final mass. This is indicative of the occurrence of delayed dynamical instability \citep{Hjellming1987} where, even if the start of the MT phase is stable, it can become unstable before the donor is fully stripped.

We thus set a hard boundary for the final Roche-lobe radius, $R_{RL,\mathrm{f}}$, of the donor at the end of the MT episode. We compute a set of \texttt{MESA} \cite[\citeyear{Paxton2013}, \citeyear{Paxton2015}, \citeyear{Paxton2018}, \citeyear{Paxton2019}]{Paxton2010} models of Helium Zero Age Main Sequence (He-ZAMS) stars with initial masses $0.275\:M_{\odot}\leq m_{\mathrm{d,f}}\leq 52.75\:M_{\odot}$. We expand the solar metallicity set of He-ZAMS in \cite{Poniatowski2021} with \texttt{MESA} v15140, and produce a set of values for the radius at He-ZAMS stage, $R_{\mathrm{He-ZAMS}}$, for each model. We then assume that the limiting orbital separation for our model to hold is the one at which the final donor's Roche-lobe radius $R_{RL,\mathrm{f}}$ is equal to the He-ZAMS radius, $R_{\mathrm{He-ZAMS}}$, of the He-ZAMS model with corresponding stripped star mass. Namely, we impose: 
\begin{equation}{\label{eq:HeMS_condition}}
R_{\mathrm{d,f}}\simeq R_{\mathrm{He-ZAMS}}<R_{RL,\mathrm{f}}\:.
\end{equation}
Physically, the condition in Eq. \ref{eq:HeMS_condition} is also implying an equilibrium configuration of our final stage after the MT episode. However, it might be an oversimplification. Investigating the details of this hard boundary requires detailed \texttt{MESA} modeling of the binary interactions and is planned in a follow-up of this work.

\subsection{Modeling $L_2$ outflow}\label{sec:L2_outflow}
\subsubsection{Shrinkage of the orbit with $L_2$ outflow}
%
Isotropic re-emission is a simplified model for the dynamics of ejected material in a binary, and real binary systems could exhibit more complex outflows. For WD accretors, high mass transfer rates ($> 10^{-6} M_{\odot}\:\mathrm{yr}^{-1}$) are expected to result in rapid expansion and the formation of a red giant star (\citealt{Nomoto1982}, \citealt{Cassisi1998}, \citealt{Shen2007}, \citealt{Wolf2013}). In a binary system this expansion will lead to overflow of outer Lagrangian points. It has also been argued that, at high mass transfer rates onto NS or BH companions, the accretion disk would grow and start ejecting most of the mass transferred through $L_2$ (\citealt{Lu2023}), preventing in some cases the expansion of the orbit after the mass ratio inversion. Such additional shrinkage sourced by $L_2$ angular momentum loss can therefore be key in favoring the formation of merging double COs, but also impact the stability of mass transfer. 

With this respect, we model the $L_2$ outflow in the form of a modified rule for the period shrinkage, $P_{\mathrm{i}}/P_{\mathrm{f}}$, using a generalized efficiency $\epsilon$ of the MT:
\begin{equation}
\epsilon =1-\beta-\upsilon\:.
\end{equation}
Here $\beta +\upsilon=0$ is the limit of conservativeness and $\upsilon =1$ corresponds to the assumption of all mass being lost through $L_2$. The $\upsilon$ efficiency is still modeled as a constant during the MT episode.

We first numerically compute the distance between the second Lagrangian point, $x_{L_2}$, and the center of mass $x_{\mathrm{cm}}\equiv 1/(1+q)$ of the system; we are calling such distance as $x_{L_2}-x_{\mathrm{cm}}$, expressed in units of the orbital separation $a$. We are considering a Cartesian system with an origin $(x=0,y=0,z=0)$ posed at the center of the donor star, with a $\hat{\vec{x}}$ axis pointing towards the accretor and $\hat{\vec{z}}$ towards the direction of the angular momentum vector. 
We then construct the following fit to the numerical solution for $(x_{L_2}-x_{\mathrm{cm}})^2$:
\begin{equation}\label{eq:L2_fit}
\left(x_{L_2}-x_{\mathrm{cm}}\right)^2(q)\simeq 1+\sum_{n=1}^{5} a_n\times
\left\{\begin{matrix}
q^{-(n-n_0)} & \hspace{0.15cm} q\geq1\\[3pt]
q^{(n-n_0)} & \hspace{0.15cm} q<1\\
\end{matrix}
\right.
\end{equation}
\[\mathrm{with}\hspace{0.25cm}n_0=0.658,\]
\[a_1=1.544,\:a_2=-3.118,\:a_3=4.387,\:a_4=-3.567,\:a_5=1.190\:.\]
The fit has a fractional error $< 0.1\%$ over the range $1\leq q<+\infty$. Thanks to the symmetry of the problem, the $L_2$ position $(x_{L_2}-x_{\mathrm{cm}})(q)$ for $q<1$ is obtained from $-(x_{L_2}-x_{\mathrm{cm}})(1/q)$. 


Material ejected through $L_2$ is assumed to carry away a specific angular momentum $j$ corresponding to this Lagrangian point, $j=a^2\:(x_{L_2}-x_{\mathrm{cm}})^2\:\Omega_{\mathrm{orb}}$, with $\Omega_{\mathrm{orb}}$ being the Keplerian angular frequency of the orbit. Following \cite{Soberman1997}, the evolution of the orbital period, $P_{\mathrm{i}}/P_{\mathrm{f}}$, as a function of the mass ratio and including both isotropic re-emission and $L_2$ mass loss can then be computed analytically as
\begin{equation}\label{eq:ratio_L2}
\dfrac{P_{\mathrm{i}}}{P_{\mathrm{f}}}=\exp\left[-3\beta\left(q_{\mathrm{f}}-q_{\mathrm{i}}\right)\right]\left(\dfrac{1+q_{\mathrm{f}}}{1+q_{\mathrm{i}}}\right)^{3\beta-1}\left(\dfrac{q_{\mathrm{f}}}{q_{\mathrm{i}}}\right)^3\times
\end{equation}
\[\exp\left[-3\upsilon \int_{q_{\mathrm{i}}}^{q_{\mathrm{f}}} f(q)\:dq\right]\hspace{0.25cm}\mathrm{with}\]
\begin{equation}\label{eq:f(q)}
f(q)\equiv \dfrac{(1+q)}{q}\:\left[x_{L_2}-x_{\mathrm{cm}}\right]^2\:.
\end{equation}
Using the fit for $(x_{L_2}-x_{\mathrm{cm}})^2$ in Eq. \ref{eq:L2_fit}, the integral $\int^q f(q)\:dq$ can be computed analytically, and we refer to App. \ref{sec:appB} for more details about its solution. 

\begin{figure}
   \includegraphics[width=0.475\textwidth]{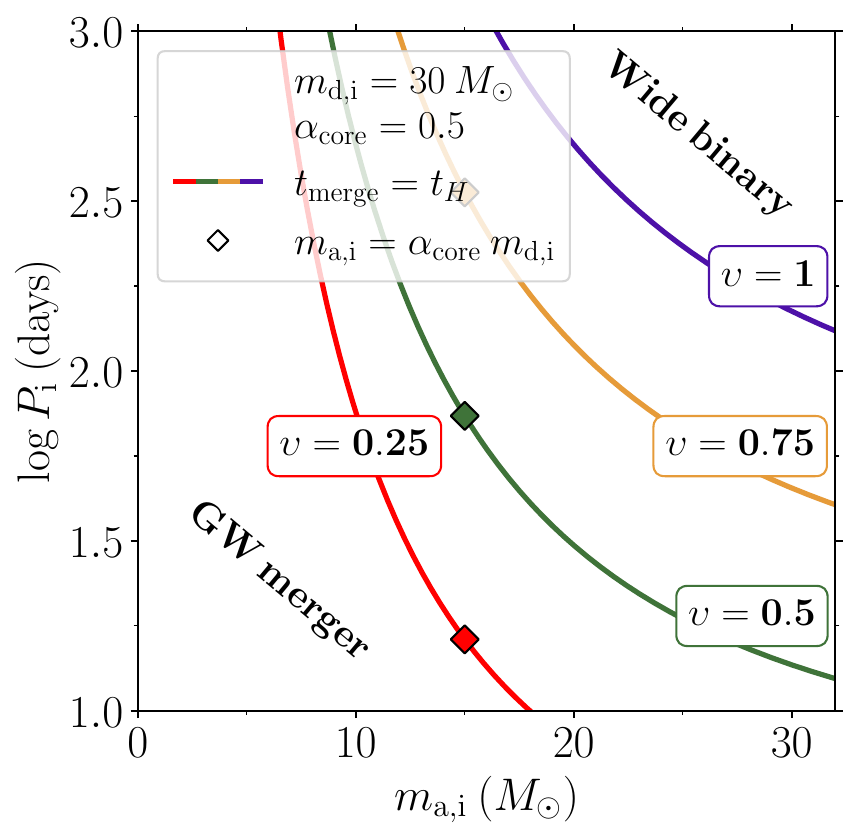}
      \caption[]{Same diagram as in Fig. \ref{fig:logPP0_first} with the modified shrinkage relation $P_{\mathrm{i}}/P$ in Eq. \ref{eq:ratio_L2}, a fixed stripping factor $\alpha_{\mathrm{core}}=0.5$ and a variable efficiency for $L_2$ mass outflow $\upsilon$. Different efficiencies $\upsilon$ are labeled along the correspondent curve.}
      {\label{fig:logPP0_L2}}
   \end{figure}

With such a modified period shrinkage rule, one can immediately define the parameter space $(m_{\mathrm{a,i}},\log P_{\mathrm{i}})$ for GW progenitor system by using Eq. \ref{eq:isotropic_reemission_condition} with the adapted $P_{\mathrm{i}}/P_{\mathrm{f}}$ of Eq. \ref{eq:ratio_L2} and switching on an efficiency for $L_2$ mass loss with $\upsilon>0$. Keeping the completely non-conservative limit $\epsilon=0$, we will then have $\beta=1-\upsilon$. The locus for fixed values of the stripping factor $\alpha_{\mathrm{core}}$ and a donor initial mass of $m_{\mathrm{d,i}}=30\:M_{\odot}$ is shown in Fig. \ref{fig:logPP0_L2}, with analogous notation as Fig. \ref{fig:logPP0_first}. As expected, we see that the extra source of angular momentum loss implies that the locus for the formation of a gravitational wave source moves towards more massive accretors. However, the expected change in orbital evolution will also impact the stability of mass transfer, as we discuss in Sect. \ref{sec:2_5_common_envelope_boundary_with_L2}. 

\subsubsection{Common envelope boundary with $L_2$ outflow}\label{sec:2_5_common_envelope_boundary_with_L2}
As an outflow from $L_2$ during the MT episode has consequences on the orbit evolution, also the orbit response to MT is affected and thus $\zeta_{RL}^{\epsilon=1-\beta}$ in Eq. \ref{eq:zeta_uno_meno_beta} needs to be properly modified:
\begin{equation}
\zeta_{RL}^{\mathrm{iso}+L_2}=\zeta_{RL}^{\epsilon=1-\beta}+2\upsilon\: qf(q)\:,
\end{equation}
and $f(q)$ is defined by Eq. \ref{eq:f(q)}. Since a $\upsilon >0$ positively shifts the value of $\zeta_{RL}^{\mathrm{iso}+L_2}$, the expected effect into the critical mass ratios $q_{\mathrm{crit}}^{\mathrm{iso}+L_2}$ is a shift towards lower values (or accordingly, more massive CO accretors) with respect to $q_{\mathrm{crit}}^{\mathrm{iso}}$. This is clearly shown in Fig. \ref{fig:zetas} for the limiting case of $\upsilon=1$. Analogously to what we did in Sect. \ref{sec:2_3_commonenvelope} for $q_{\mathrm{crit}}^{\mathrm{iso}}$, we determine $q_{\mathrm{crit}}^{\mathrm{iso}+L_2}$ for a donor with a given initial mass $m_{\mathrm{d.i}}$ and radius $R_{\mathrm{d,i}}$, in presence of $\upsilon >0$ and $\epsilon=0$ by imposing the condition:
\begin{equation}\label{eq:stability_L2_condition}
\zeta_{\mathrm{ad}}(R_{\mathrm{d,i}},m_{\mathrm{d,i}})=\zeta_{RL}^{\mathrm{iso}+L_2}(q_{\mathrm{crit}}^{\mathrm{iso}+L_2})\:.
\end{equation}

\section{Results}\label{sec:Results}
This section presents the results about the favorable parameter space for GW progenitor systems, in the form of $(m_{\mathrm{a,i}},\log P_{\mathrm{i}})$ diagrams with fixed donor mass $m_{\mathrm{d,i}}$. We show the results for a BH direct collapse (Sect. \ref{sec:BHdonor}), NS remnant with kick velocity (Sect. \ref{sec:NSdonor}) and WD accretors (Sect. \ref{sec:WDdonor}); just one example for fixed $m_{\mathrm{d,i}}$ is shown for each case. The full set of results for $2<m_{\mathrm{d,i}}/M_{\odot}<100$ will be made publicly available at the time of publication on Zenodo\footnote{\href{doi.org/10.5281/zenodo.8334056}{doi.org/10.5281/zenodo.8334056}}.

\subsection{BH progenitor as the donor}\label{sec:BHdonor}
When the initial mass $m_{\mathrm{d,i}}$ of the donor is $m_{\mathrm{d,i}}\geq 25\:M_{\odot}$, we assume that, after a later evolution of the stripped star following the MT episode, the star undergoes direct collapse into a BH. As such, we model the evolutionary history from the end of the MT episode until merging as if the direct collapse happens immediately after the MT. Therefore, this model ignores possible mass loss between the end of the MT phase and BH formation, which is expected to widen the orbit (see discussion in Sect. \ref{sec:Effect of winds}). We briefly discuss the possibility of having BHs kick for $25\lesssim m_{\mathrm{d,i}}\:(M_{\odot})\lesssim 40$ in Sec. \ref{sec:NSdonor}. Afterwards, the double CO system evolves only under the effect of gravitational interaction via quadrupole radiation emission, see Eq. \ref{eq:Peters}.

\begin{figure}
   \centering
   \includegraphics[width=0.4725\textwidth,scale=0.8]{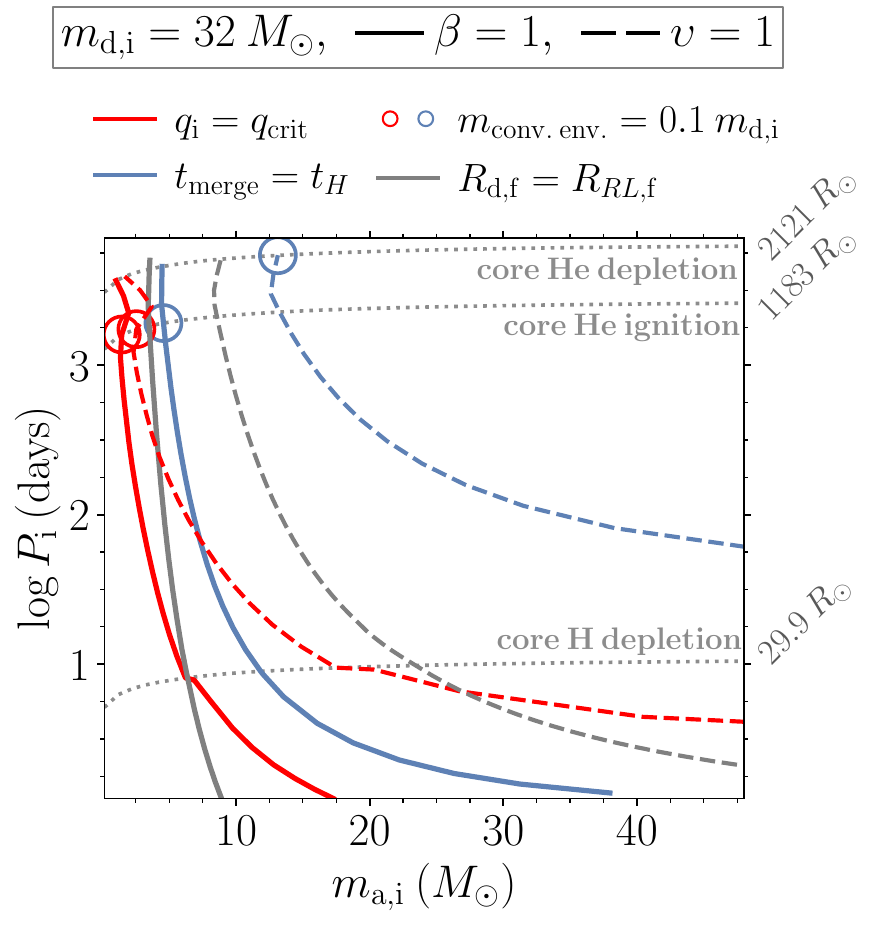}
      \caption[]{Boundaries for unstable MT (red) and formation of GW sources (blue) for a 32 $M_\odot$ donor orbiting a CO of mass $m_\mathrm{a,i}$. The area below the gray line excludes systems for which the final orbital separation implies $R_\mathrm{d,f}<R_\mathrm{RL,f}$. Results are shown for full isotropic re-emission (solid), and assuming all matter is ejected from $L_2$ (dashed). Different stages of evolution are illustrated with contours of constant Roche lobe radius (dotted gray). Empty circles mark the formation of a convective envelope containing $10\%$ of the donor's mass.}
      {\label{fig:mdonor32_L2}}
   \end{figure}

Figure \ref{fig:mdonor32_L2} shows the parameter space $(m_{\mathrm{a,i}},\log P_{\mathrm{i}})$ for GW progenitor systems as results from the described approach. The Roche lobe filling star is a $m_{\mathrm{d,i}}=32\:M_{\odot}$ donor.
Let us first consider the purely isotropic re-emission channel (solid lines). For any orbital period there is a range in mass ratio where MT is expected to be stable and lead to sufficient hardening to produce a CO coalescence within a Hubble time. Such a strip extends between the common envelope boundary (solid red) and the delay time boundary (solid blue). Requiring that $R_{RL,\mathrm{f}}>R_{\mathrm{d,f}}$ at the end of the MT episode, as discussed in Sect. \ref{sec:Final_orbital_separation}, limits significantly such strip for more evolved donors. Nevertheless, the accretor masses $m_{\mathrm{a,f}}$ spanned by the strip, at different stages, are such that $5\lesssim m_{\mathrm{a,i}}/M_{\odot}\lesssim 37$. Therefore, we confirm that the production of merging double BH from stable MT via isotropic re-emission is possible. This, however, relies on the assumption that such initial conditions are realized in nature, which we can inform using observations of single-degenerate binaries (see Sect. \ref{sec:3_4_comparison_known_systems}). We also rely on the assumption that the response to MT of the donor star, which potentially accreted mass from the BH progenitor in a previous evolutionary phase, is well approximated with the single star models of \cite{Ge2020}.

In Fig. \ref{fig:mdonor32_L2} we also show the results after switching on an efficiency $\upsilon=1$ for $L_2$ mass outflow. As illustrated in the simple model with fixed $\alpha_\mathrm{core}$ of Fig. \ref{fig:logPP0_L2}, the locus for the formation of a merging compact object is shifted towards higher accretor masses. However, the additional angular momentum loss also shifts the boundary for stability towards higher accretor masses. Nevertheless, a range in mass ratio and initial periods for merging double COs, from stable MT for $\upsilon>0$, is still predicted. The masses for the compact object accretor in this strip are such that $m_{\mathrm{a,i}} (M_{\odot})\gtrsim 11$, thus the expected outcome is indeed merging binary black holes, though with different mass ratios with respect to the purely isotropic re-emission case.

The predicted strip for $m_{\mathrm{d,i}}=32\:M_{\odot}$ can be compared to the one derived, with a set of \texttt{MESA} simulations, by \cite{Marchant2021} for a 30 $M_{\odot}$ metal poor ($Z=0.00142$), Roche lobe filling donor star orbiting a CO. The results from our semi-analytical model are qualitatively consistent with the ones from their set of detailed simulations, for both the instability boundary and the merging time constraint, although our predictions overestimate the set of initial conditions that lead to a GW source, see App. \ref{app:Marchant_comparison} and Fig. \ref{fig:PabloCOMP} for a discussion. As discussed in App. \ref{app:Marchant_comparison}, this holds for our results in the purely isotropic re-emission channel, as in their simulations the accretion rate for the stable MT episodes easily approaches the Eddington one, so that $\beta\simeq 1$ even though in the simulations the degree of non-conservativeness is not a free parameter.

\begin{figure}
   \centering
   \includegraphics[width=0.4725\textwidth,scale=0.8]{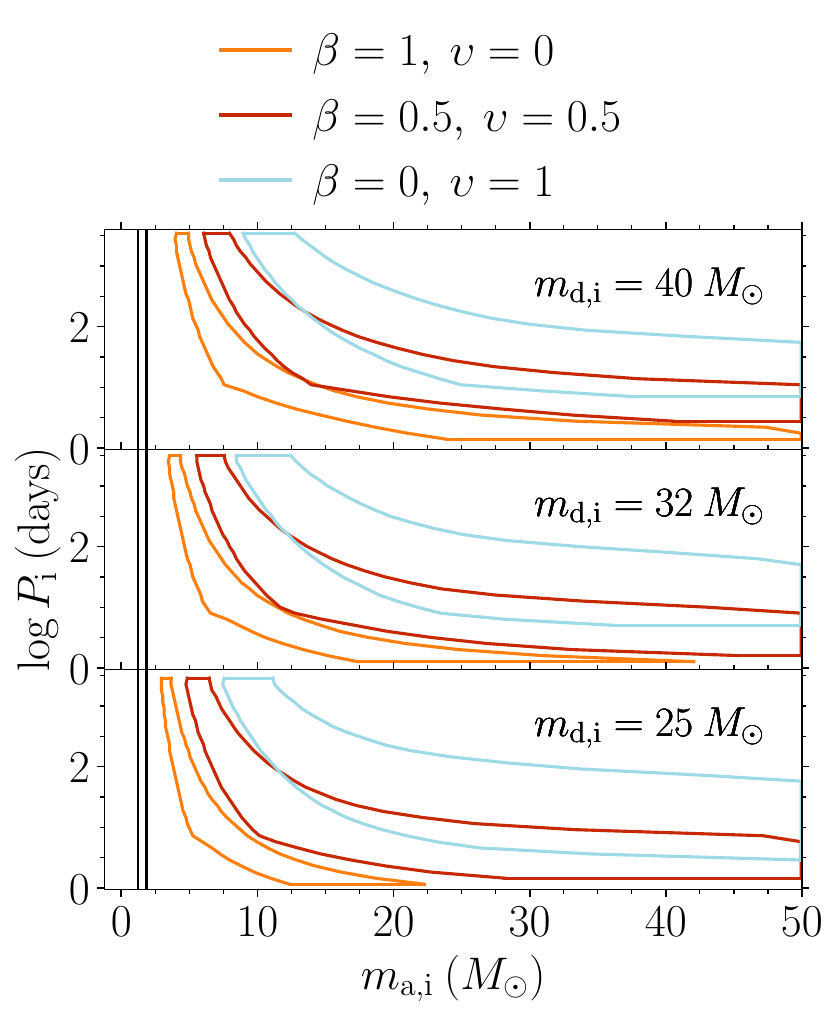}
      \caption[]{Locus for the formation of a GW source through stable mass transfer with $40\:M_{\odot},\:32\:M_{\odot},\:25\:M_{\odot}$ donors, i.e. BH progenitors, as a function of the CO accretor mass $m_\mathrm{a,i}$ and orbital period $P_\mathrm{i}$. Different colors mark the favorable strips for merging double compact objects via purely isotropic re-emission ($\beta=1$) and mass outflow from $L_2$ with variable efficiencies $\upsilon>0$, keeping $\beta+\upsilon=1$. Black solid lines mark the range of known NS masses from binary systems (\citealt{Tauris2017}).}
      {\label{fig:contours_L2_BHs}}
   \end{figure}

Figure \ref{fig:contours_L2_BHs} summarizes our results for the GW merger strip, in the form of a $(m_{\mathrm{a,i}},\log P_{\mathrm{i}})$ diagram similar to Fig. \ref{fig:mdonor32_L2}, with different donor masses in the range of BH progenitors, i.e. $m_{\mathrm{d,i}}=40\:M_{\odot},\:32\:M_{\odot},\:25\:M_{\odot}$. In these diagrams we vary discretely the efficiencies for isotropic re-emission and mass outflow from $L_2$, keeping the total $\epsilon=1-\beta-\upsilon=0$. Firstly, we can see a shift towards larger accretor masses (i.e., smaller final mass ratios $q_{\mathrm{f}}$) as we increase $m_{\mathrm{d,i}}$. Secondly, we notice the possibility, for the second born CO, to result in either the more massive object of the merging CO + CO pair (this happens at the longer initial periods) \textit{or} the less massive one (lower initial periods). Thirdly, we observe that the coupling of the second born BH with a first born NS accretor, within the stable MT channel, is excluded.


Although we place no constraints on the first born CO mass in our analysis, stellar evolution can prevent the formation of arbitrarily large BH masses $m_{\mathrm{a, i}}\gtrsim 30\:M_{\odot}$ as considered in Fig. \ref{fig:contours_L2_BHs}. At the $Z=0.02$ metallicity of the \citealt{Ge2020} simulation set, stellar winds potentially prevent the formation of such massive BHs (e.g., \citealt{Spera2015}). Nevertheless, we do not restrict the BH masses we study as we take the \citealt{Ge2020} results to be also representative of lower metallicity environments, which is supported by our comparison to the stability boundary at low metallicity of \citealt{Marchant2021}, Appendix \ref{app:Marchant_comparison}.


\subsection{NS progenitor as the donor}\label{sec:NSdonor}
The range of ZAMS masses $m_{\mathrm{ZAMS}}$ for leaving behind a SN remnant is a matter of debate: the upper ZAMS mass limit, classically posed at $m_{\mathrm{ZAMS}}\lesssim 25\:M_{\odot}$ by, e.g., \cite{Fryer1999}, has been recently argued to be formed by “islands of explodability” (e.g. \citealt{O'Connor2011}, \citealt{Ugliano2012}, \citealt{Muller2016}, \citealt{Mabanta2019}). On the other hand, the lower ZAMS masses limit is canonically set at $m_{\mathrm{ZAMS}}\gtrsim 8\:M_{\odot}$ based on single massive stars evolution studies (e.g. \citealt{Poelarends2008}), but depends on metallicity (e.g. \citealt{Langer2012}) and can be different for stars born in close binaries (\citealt{Wellstein2001}). In the following, we ignore the above complications and adopt the simplistic range $8< m_{\mathrm{ZAMS}}/M_{\odot}< 25$. In particular, when the initial mass $m_{\mathrm{d,i}}$ of the donor is such that $8\leq m_{\mathrm{d,i}}/M_{\odot}< 25$, we assume that, after the MT episode and the later evolution of the stripped star, it explodes as a supernova (SN) leaving a NS compact remnant.

Since the population of galactic NSs is known to have a velocity dispersion much larger than that of OB-type stars, indicative of kicks during a SN explosion (see \citealt{Lai2001} for a discussion), we also take into account this possibility in our calculations. As also BHs have been argued to have potential kick velocities at birth (see e.g. \citealt{Atri2019}), we also considered the case of a $m_{\mathrm{d,i}}=32\:M_{\odot}$ BH progenitor at the end of this section. 

For the case of NSs, the post-SN orbital parameters have been derived by several authors (e.g. \citealt{Sutantyo1978}, \citealt{Hills1983}, \citealt{Verbunt1990}, \citealt{Brandt1995}, \citealt{Kalogera1996}); in particular, we will follow the notation and expressions given in \cite{Tauris1999} (see equations 6 - 8 of that work). We point out that the calculation of the post-SN properties given by \cite{Tauris1999} assumes a circular pre-SN orbit followed by an instantaneous SN explosion and subsequent kick. 

With these same assumptions, and adopting a canonical mass of $m_{\mathrm{d,\:pSN}}=1.4\:M_{\odot}$ for the second born CO, i.e. the NS remnant, we calculate the orbital separation $a_{\mathrm{pSN}}$ and the acquired eccentricity $e_{\mathrm{pSN}}$ of the system after the SN explosion; we used the subscript $\mathrm{pSN}$ to indicate the post-SN stage, to differentiate with the subscript $\mathrm{f}$ which we used above for the post-MT configuration. Also, as opposed to \cite{Tauris1999}, we do not take into account the possible effects, on the orbit and on the first born compact companion, of the passing shock wave from the SN (i.e., we set $v_{\mathrm{im}}=0$ in \citealt{Tauris1999}).


To assess if the binary system at the end of the MT can evolve into a GW source, we adopt the analytical expression for the delay time $t_{\mathrm{merge}}$ found by \cite{Peters1964}:
\begin{equation}\label{eq:Peters_eccentric}
t_{\mathrm{merge}}=\dfrac{12}{19}\dfrac{c_{\mathrm{pSN}}^4}{\mathcal{B}}\int_0^{e_{\mathrm{pSN}}}de\: \dfrac{e^{29/19}[1+(121/304)\:e^2]^{1181/2299}}{(1-e^2)^{3/2}}\:,
\end{equation}
\begin{equation}
\text{with}\hspace{0.5cm} c_{\mathrm{pSN}}^4\equiv a_{\mathrm{pSN}}^4\left[\dfrac{1-e_{\mathrm{pSN}}^2}{e_{\mathrm{pSN}}^{12/19}}\right]^4\left\{\left[1+\dfrac{121}{304}e_{\mathrm{pSN}}^2\right]^{-870/2299}\right\}^4\:.
\end{equation}
This expression is the generalized form of Eq. \ref{eq:Peters} in the case of non-zero post-SN eccentricity.

To account for the possibility of kicks, we adopt the kick velocity distribution in \cite{Hobbs2005} for the second born CO, i.e. a 1D Maxwellian with dispersion $260\:\mathrm{km}\:\mathrm{s}^{-1}$, and isotropic orientation in a frame of reference comoving with the SNe progenitor. For each initial orbital period $P_{\mathrm{i}}$ and compact object accretor mass $m_{\mathrm{a,i}}$, we compute the post-MT orbital properties in the same way as we did for the BH progenitor in Sect. \ref{sec:BHdonor}, obtaining a distribution of the post-SN properties, i.e. $e_{\mathrm{pSN}},\: a_{\mathrm{pSN}}$ due to the post-SN kick. Because of this, for each set of donor mass $8\leq m_{\mathrm{d,i}}/M_{\odot}< 25$, accretor mass and initial orbital period, we have a probability $p_{t_{\mathrm{merge}}<t_H}$ of forming a merging compact object binary.

\begin{figure}
   \centering
   \includegraphics[width=0.475\textwidth]{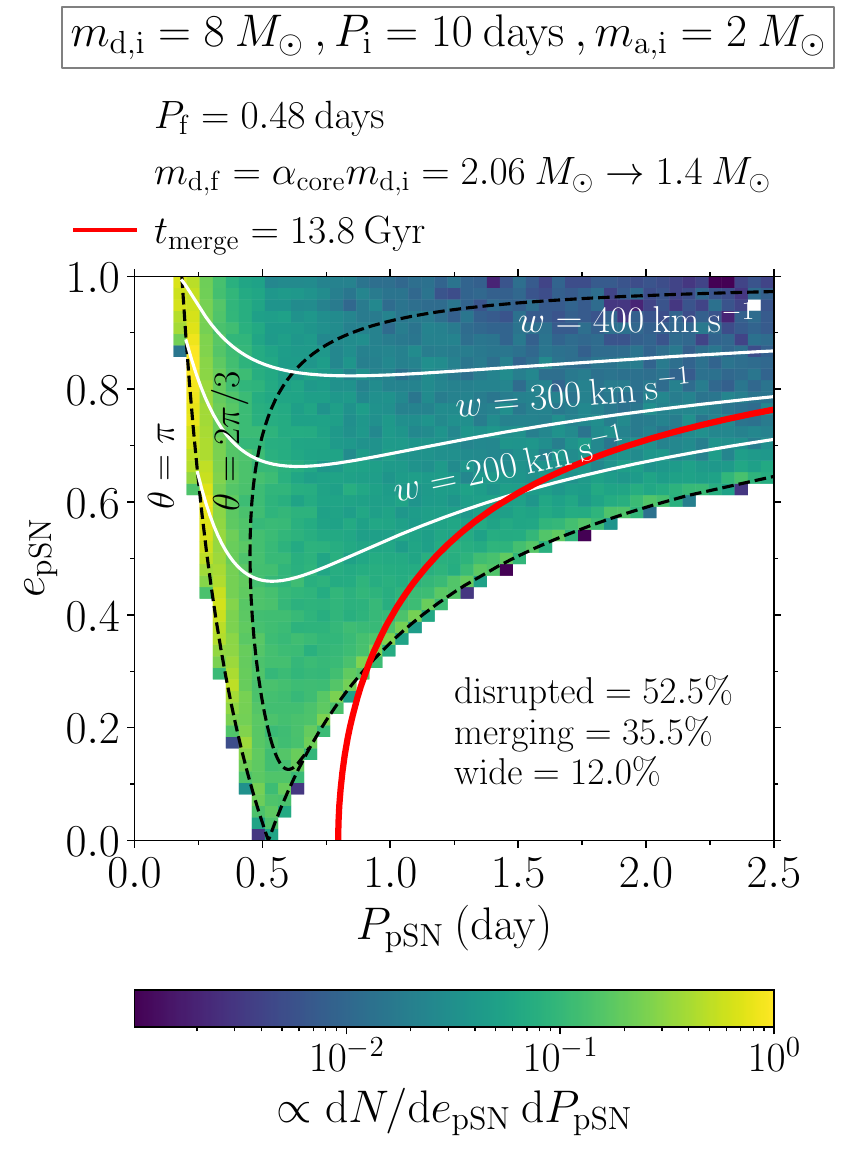}
      \caption{Distribution of post-SN eccentricity $e_{\mathrm{pSN}}$ and orbital period $P_{\mathrm{pSN}}$ for an example binary system (see Sect. \ref{sec:NSdonor}  for details). Solid white lines indicate final orbital parameters for fixed kick velocity $w$ and variable angle $\theta$ formed between the kick velocity and the orbital velocity, assuming the kick to be on the orbital plane (i.e. $\phi$ = 0). Dashed black lines indicate the same for a fixed $\theta$ and a variable $w$. A thick solid red line shows the boundary separating wide (right of the curve) from merging (left of the curve) systems. }
      {\label{fig:postSN_properties} }
   \end{figure}

To illustrate the impact of a SN kick, we consider one specific binary system with an $8\:M_\odot$ donor star and a $2\:M_\odot$ accretor with an orbital period of $10$ days. The binary undergoes a stable MT episode with isotropic re-emission, after which the donor star is stripped to $m_{\mathrm{d,f}}=2.06\:M_{\odot}$ and the orbital period is $P_{\mathrm{f}}=0.48\:\mathrm{days}$, and these are considered as the pre-SN properties. We consider the occurrence of a kick to the resulting compact object according to the distribution of kick properties described in the previous paragraph. There is a 52.5\% probability that the binary is disrupted from the kick. For the remaining 47.5\%, Fig. \ref{fig:postSN_properties} illustrates the resulting distribution of post-SN eccentricities and periods: a fraction 35.5\% of them has conditions that allow for a merger within 13.8 Gyr, and the remaining 12\% end up being too wide for GW radiation to drive to a merger within the age of the Universe.


\begin{figure}
   \centering
   \includegraphics[width=0.475\textwidth]{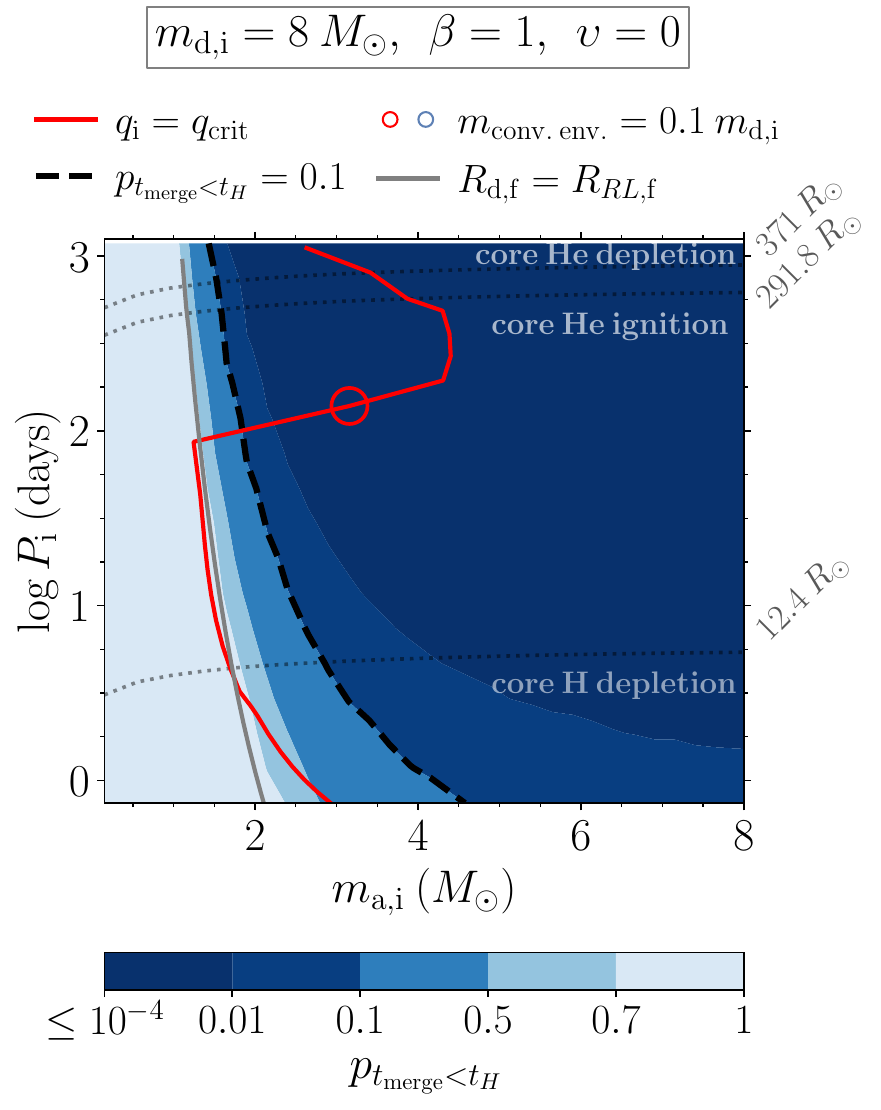}
      \caption[]
      {\label{fig:mdonor8_L2} Probability to form a GW source through stable mass transfer with an $8\:M_\odot$ donor as a function of the CO accretor mass $m_\mathrm{a,i}$ and orbital period $P_\mathrm{i}$. The donor star is assumed to form a neutron star with an isotropic kick (see Sect. \ref{sec:NSdonor}).  Analogous symbols and notation as in Fig. \ref{fig:mdonor32_L2} are employed for the purely isotropic re-emission channel ($\beta=1$). The contour color-code maps probabilities $p_{t_{\mathrm{merge}}<t_{H}}$ of obtaining merging times $t_{\mathrm{merge}}<t_H$ given the distribution of kick velocities from \cite{Hobbs2005} and isotropic kick directions, as explained in Sect. \ref{sec:NSdonor}.}
   \end{figure}

\begin{figure}
   \centering
   \includegraphics[width=0.475\textwidth]{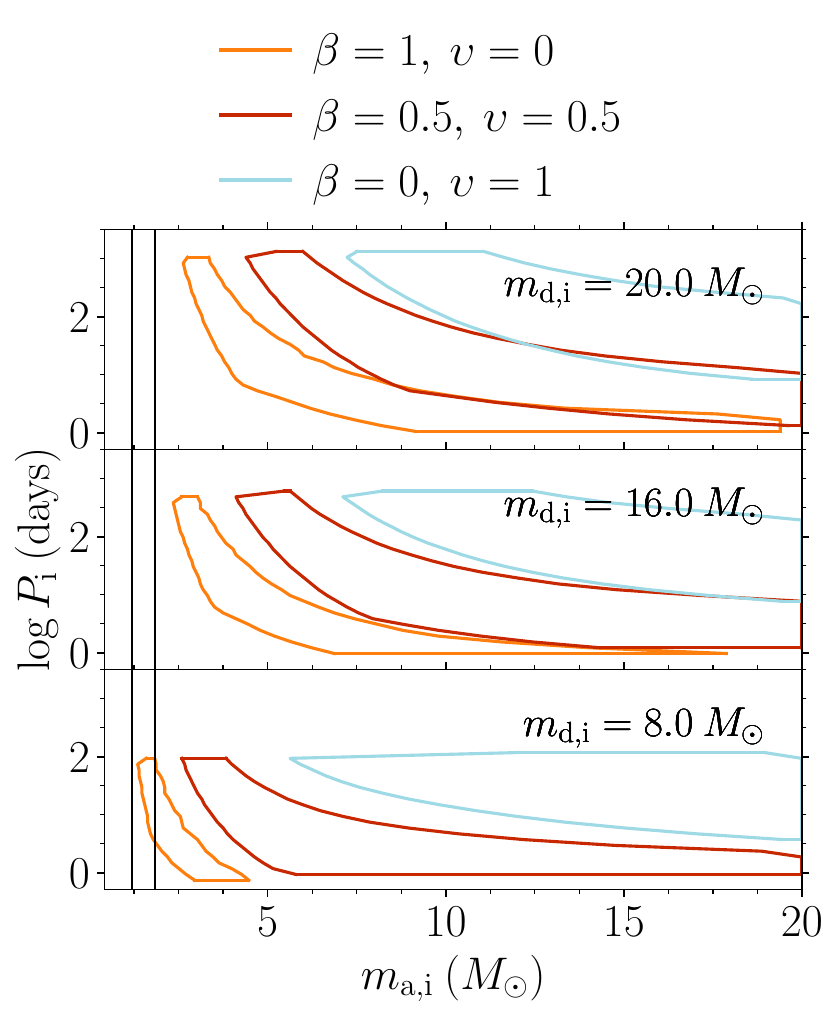}
      \caption[]{Locus for the formation of a GW source through stable mass transfer with $20\:M_{\odot},\:16\:M_{\odot},\:8\:M_{\odot}$ donor, i.e. NS objects progenitors, as a function of the CO accretor mass $m_\mathrm{a,i}$ and orbital period $P_\mathrm{i}$. Contours indicate the region where the merger probability after the supernova kick is $p_{t_{\mathrm{merge}}<t_H}>0.1$. Analogous notation to Fig. \ref{fig:contours_L2_BHs} is employed. }
      {\label{fig:contours_L2_NSs}}
   \end{figure}

The resulting merger probabilities are shown in Fig. \ref{fig:mdonor8_L2} for the case of a donor star of $m_\mathrm{d,i}=8\:M_\odot$ under the assumption of isotropic re-emission. The favorable strip for GW mergers comprises a small interval of accretor masses, and for later evolutionary stages the common envelope boundary (solid red) takes over and overlays any possible system with $t_{\mathrm{merge}}<t_H$. For this 8 $M_{\odot}$ donor, we find that the compact object accretors for which we form a GW source span, at different evolutionary stages, between $1.75\lesssim m_{\mathrm{a,i}}(M_{\odot})\lesssim 3.75$. As we can observe in Fig. \ref{fig:contours_L2_NSs} looking at the $\beta=1$ contours, for the case of a $m_{\mathrm{d,i}}=16\:M_{\odot}$ and a $m_{\mathrm{d,i}}=20\:M_{\odot}$ donor we find that the range of CO accretors spans $2\lesssim m_{\mathrm{a,i}}(M_{\odot})\lesssim 10$ and $3\lesssim m_{\mathrm{a,i}}(M_{\odot})\lesssim 15$, respectively. For a narrow range of donor masses, just above the limit at which we expect core-collapse SNe, stable MT can then produce merging binary neutron stars. However, for a wider range of donor masses that would result in a NS, we require a compact object accretor with a mass consistent with a BH.


The bottom panel of Fig. \ref{fig:contours_L2_NSs} shows the results for $m_{\mathrm{d,i}}=8\:M_{\odot}$ after switching on an efficiency $\upsilon=1$ for $L_2$ mass outflow (light blue area). The pairing between NS and BH is here predicted all along the favorable GW strip. However, the additional angular momentum loss during the MT episode removes the possibility of forming a merging binary neutron star, allowing solely for the formation of merging NS + BH systems where the NS is the second born CO. The other panels in Fig. \ref{fig:contours_L2_NSs} show the results for $m_{\mathrm{d,i}}=20\:M_{\odot},\:16\:M_{\odot}$ and different efficiencies $\beta$ and $\upsilon$. For these cases, we confirm the aforesaid prediction on merging NS + BH pairs with a second born NS; this is true for any value of $L_2$ outflow efficiency $\upsilon>0$. 

Lastly, to investigate on the effect of a possible kick imparted to a BH progenitor, we calculated the probability $p_{t_{\mathrm{merge}}<t_H}$, in the same fashion as done for NS progenitors, for the case $m_{\mathrm{d,i}}=32\:M_{\odot}$. We assumed a) mass loss of $10\%$, i.e. the mass of the second born BH is assumed to be $m_{\mathrm{d,\:pSN}}=0.1\times m_{\mathrm{d,f}}$; b) an imparted kick magnitude of $100\:\mathrm{km}\:\mathrm{s}^{-1}$. The resulting parameter space for GW mergers is superimposable with the direct collapse results in Fig. \ref{fig:mdonor32_L2}. Assuming a small, fixed fraction of mass loss makes the system stay massive, thus $p_{t_{\mathrm{merge}}<t_H}\simeq 1$ are found below the time delay boundary at any $\log P_{\mathrm{i}}$.

\subsection{WD progenitor as the donor}\label{sec:WDdonor}
When the initial mass of the donor is $m_{\mathrm{d,i}}<8\:M_{\odot}$ we assume that, after the MT episode, the stripped star evolves into a WD, orbiting its compact companion with no further interaction.

\begin{figure}
   \centering
   \includegraphics[width=0.475\textwidth]{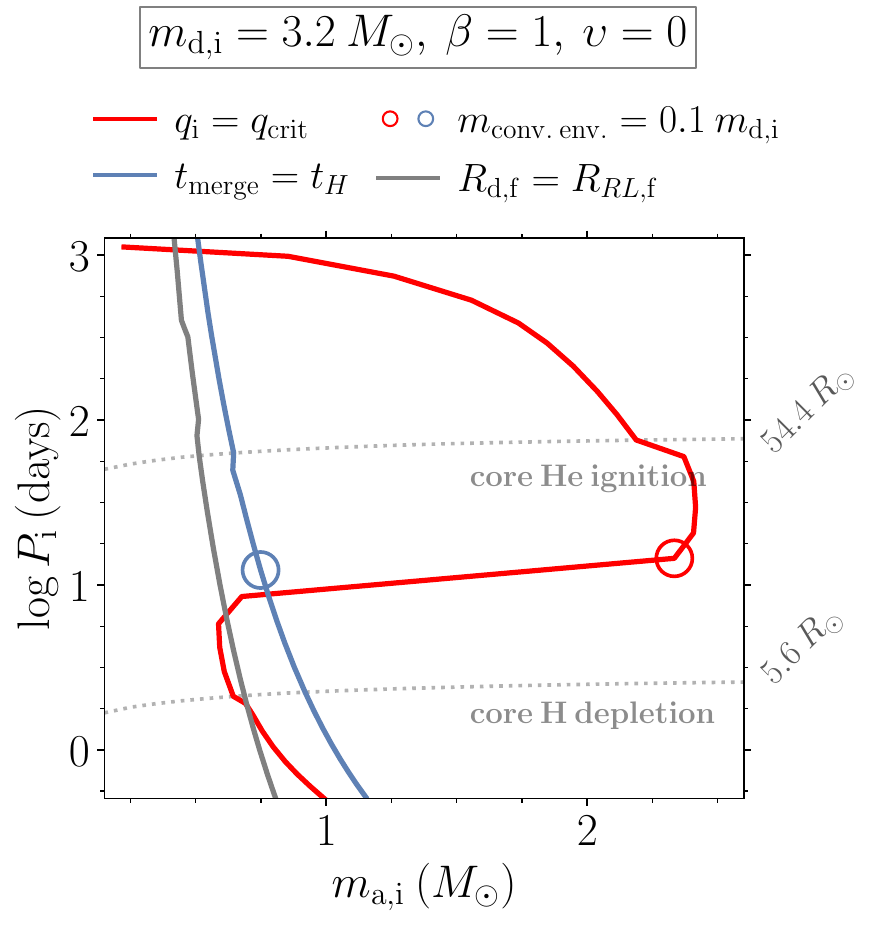}
      \caption[]{Boundaries for unstable MT (red) and formation of GW sources (blue) for a 3.2 $M_\odot$ donor orbiting a CO of mass $m_{\mathrm{a,i}}$. Analogous symbols and notation as in Fig. \ref{fig:mdonor32_L2} are employed for the purely isotropic re-emission channel $\beta=1$.}
      {\label{fig:mdonor3_2_L2}}
   \end{figure}

\begin{figure}
   \centering
   \includegraphics[width=0.475\textwidth]{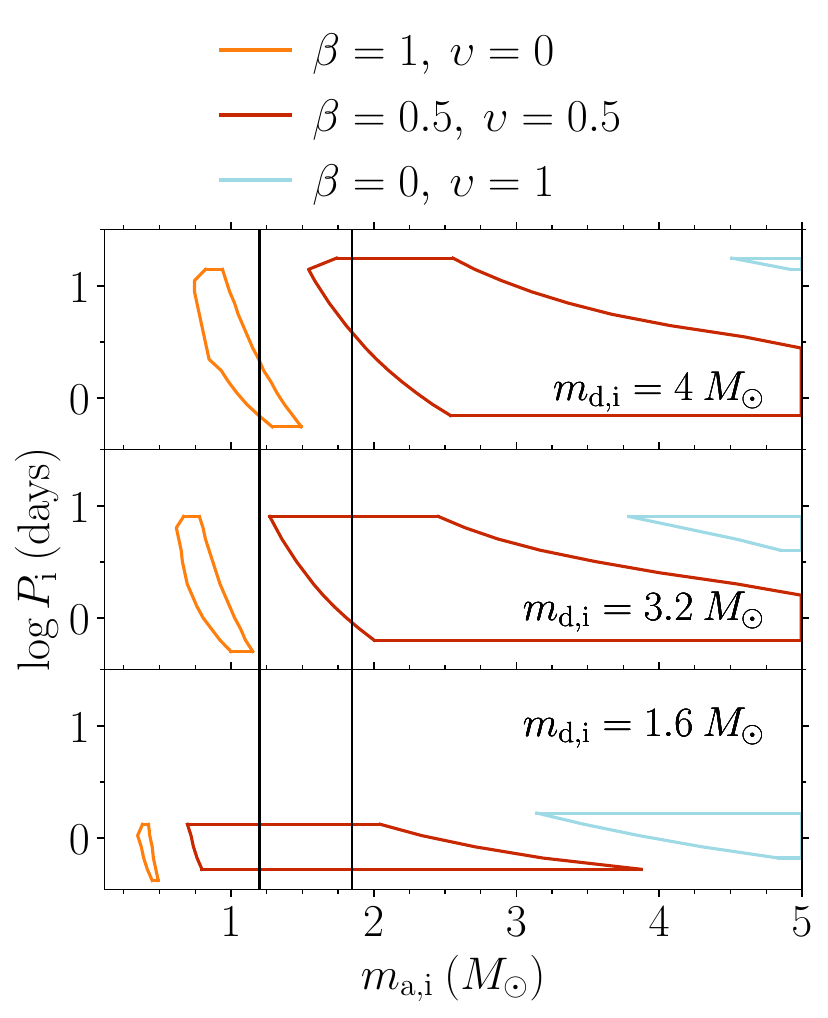}
      \caption[]{Locus for the formation of a GW source through stable mass transfer with $4\:M_{\odot},\:3.2\:M_{\odot},\:1.6\:M_{\odot}$ donors, i.e. WD objects progenitors, as a function of the CO accretor mass $m_\mathrm{a,i}$ and orbital period $P_\mathrm{i}$. Analogous notation to Fig. \ref{fig:contours_L2_BHs} is employed.}
      {\label{fig:contours_L2_WDs}}
   \end{figure}

Figure \ref{fig:mdonor3_2_L2} shows the resulting diagram $(m_{\mathrm{a,i}},\log P_{\mathrm{i}})$ for GW progenitors for the case of $m_{\mathrm{d,i}}=3.2\:M_{\odot}$. In the purely isotropic re-emission channel (solid lines), the favorable parameter space for merging WD + CO from stable MT is visibly limited to a thin strip of mostly systems with pre-TAMS interaction. A small post-TAMS region is also comprised between the common envelope (solid red) and the merging time (solid blue) boundaries. The possible accretor masses $m_{\mathrm{a,i}}$ at different initial orbital period are such that $0.75\lesssim m_{\mathrm{a,i}}/M_{\odot}\lesssim 1.25$: this predicts merging WD + WD pairs. As shown in Fig. \ref{fig:contours_L2_WDs} for the $\beta=1$ contours, when we scale the donor mass $m_{\mathrm{d,i}}$ to larger values the strip shifts to larger values for $m_{\mathrm{a,i}}$. For $m_{\mathrm{d,i}}=4\:M_{\odot}$ in Fig. \ref{fig:contours_L2_WDs} (top panel), we find that the $m_{\mathrm{a,i}}$ accretor masses spanned are such that $0.8\lesssim m_{\mathrm{a,i}}/M_{\odot}\lesssim 1.5$. We thus find space for a merging system with a first born NS orbiting the second born WD. On the other hand, for $m_{\mathrm{d,i}}=2\:M_{\odot}$ in Fig. \ref{fig:contours_L2_WDs} (bottom panel), we see that the favorable region is almost entirely suppressed.

The middle panel of Fig. \ref{fig:contours_L2_WDs} shows the results for $m_{\mathrm{d,i}}=3.2\:M_{\odot}$ with an efficiency $\upsilon=1$ for $L_2$ mass outflow (light blue area). Similar to the BH and the NS progenitor cases, the strip to form a GW source via stable MT is shifted towards larger accretor masses with respect to the purely isotropic re-emission case. In this case, it could also be consistent with the formation of a WD + BH compact binary, where the WD is the second born CO. This is also shown in the other panels of Fig. \ref{fig:contours_L2_WDs} with the results for $m_{\mathrm{d,i}}=4\:M_{\odot},\:1.6\:M_{\odot}$: the additional angular momentum loss ($\upsilon>0.5$ contours) allows for the formation of the WD + BH coupling for each of these systems. 
Like for the earlier cases, our results depend on whether systems with these properties form in nature, as the realization of such initial conditions is sensitive to the evolution of the binary system prior to the formation of the first born CO.


\subsection{The effect of wind mass loss}\label{sec:Effect of winds}
The set of simulations from Ge et al. series \cite[\citeyear{Ge2015}, \citeyear{Ge2010}]{Ge2020} has been produced without modeling wind mass loss. We may expect, however, that mass lost via winds from the donor stars could impact their radii $R_{\mathrm{d,i}}$, thus the evolutionary stage at which the MT episode is initiated, as well as the core mass $m_{\mathrm{c}}$, hence the post-MT stage of the systems. This is especially true for the larger masses of the grid, $m_{\mathrm{d,i}}\gtrsim 10\:M_{\odot}$, for which large fractions of the initial mass might be expelled via radiation-driven winds in a $Z=0.02$ environment. 

Wind mass loss from the more massive donor stars can also impact the orbital evolution between the end of the MT episode and the formation of the second born CO. Mass loss after MT translates into larger merging times, as the final CO + CO system is less massive when gravitational radiation takes over in determining the orbital evolution during the inspiral phase. Additionally, mass lost from the vicinity of the donor star with its specific angular momentum is expected to widen the orbit. Angular momentum balance arguments yield that the orbital period $P_{\mathrm{f}}$ at the end of the MT episode can be widened to $P_{\mathrm{f,wind}}$ by a correcting factor
\begin{equation}\label{eq:WIND_ML_ratio}
\dfrac{P_{\mathrm{f,wind}}}{P_{\mathrm{f}}}=\left(\dfrac{q_{\mathrm{f}}+1}{q_{\mathrm{f,wind}}+1}\right)^2\:>1\:,
\end{equation}
which solely depends on the mass ratios $q_{\mathrm{f}}=m_{\mathrm{d,f}}/m_{\mathrm{a,f}}$ and $q_{\mathrm{f,wind}}\equiv m_{\mathrm{d,f}}|_{\mathrm{wind}}/m_{\mathrm{a,f}}$. We are here referring to a $P_{\mathrm{f,wind}}$ period at which the donor star has reached the mass $m_{\mathrm{d,f}}|_{\mathrm{wind}}$ due to the wind mass loss effect.

As this correction can have an impact in resolving the GW mergers boundaries, we want to be more quantitative and study the impact of wind mass loss in systems composed by a CO orbiting a donor star of mass $m_{\mathrm{d,i}}=32\: M_{\odot}$. To do so, we computed a set of simulations with \texttt{MESA}. We model the evolution of helium stars with initial masses $m_{\mathrm{d,f}}$ from \cite{Ge2020} (see Sect. \ref{sec:2_4_simulations}) resulting after the stripping MT episode from the donor star with $m_{\mathrm{d,i}}=32\: M_{\odot}$. The evolution of the models set is stopped at the end of the stars' core C-burning phase. During this evolution, wind mass loss following the prescription from \cite{SanderVink2020} is computed and accounted for. We use the final masses $m_{\mathrm{d,f}}|_{\mathrm{wind}}$ from the simulations to re-compute the merging times $t_{\mathrm{merge}}|_{\mathrm{wind}}$ with the described corrections on the masses and the orbital separation at the start of the inspiral phase. 

\begin{figure}
   \resizebox{\hsize}{!}{
   \includegraphics[width=1.\textwidth]{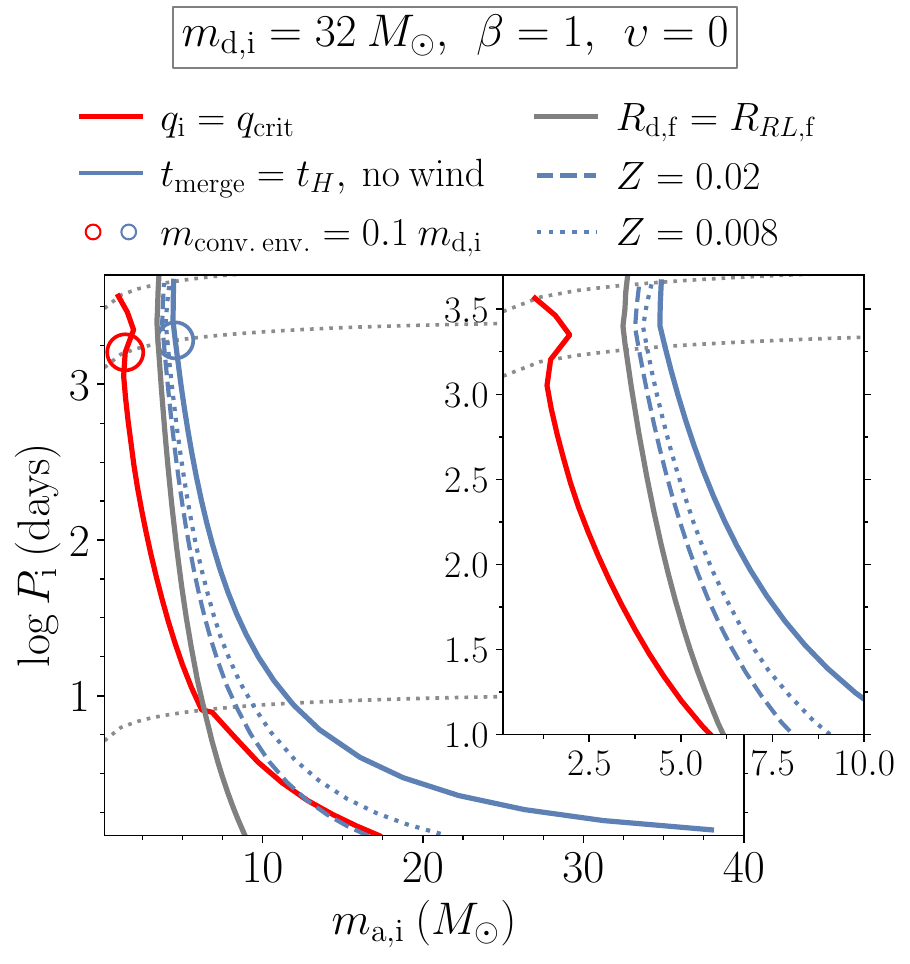}}
    \caption[]{{\label{fig:WIND_ML_32}} Same as Fig. \ref{fig:mdonor32_L2}, but illustrating the impact of wind mass loss after the phase of stable MT. The dashed (dotted) blue line shows the modified delay time boundary (solid blue) due to wind mass loss effects, in an environment with $Z=0.02$ ($Z=0.008$), on the post-MT mass of the stripped star and on the orbital separation at the start of the inspiral phase, see text in Sect. \ref{sec:Effect of winds} for more details. Notice that we are considering the effect under the isotropic re-emission $\beta=1$ assumption on the MT episode. For the rest, analogous symbols and notation as in Fig. \ref{fig:mdonor32_L2} are employed.}
   \end{figure}

We computed the correction in Eq. \ref{eq:WIND_ML_ratio} first with a metallicity $Z=0.02$, suitable for a Milky Way-like environment; subsequently, we explored the possibility of a Large Magellanic Cloud-like environment ($Z=0.008$). Figure \ref{fig:WIND_ML_32} shows the resulting modified boundaries for the isotropic re-emission channel, both for the Milky Way and for the Large Magellanic Cloud. The favorable strip for merging double COs is significantly reduced, as the time delay boundary is shifted towards larger initial mass ratios and eventually reaches the Roche-lobe boundary at advanced evolutionary stages. This is especially true for $Z=0.02$.

This test-case shows that the effect of wind mass loss physics on our results can be impactful, at least in the cases of massive donors ($m_{\mathrm{d,i}}\gtrsim 10\: M_{\odot}$) for Milky Way-like environments. We thus reserve a more extensive investigation of winds to future work. 

\subsection{Including conservativeness $\epsilon>0$}\label{sec:Conservativeness}
So far we have carried out our calculations assuming completely non-conservative MT, either in the isotropic re-emission channel with $\beta=1$ and $\epsilon=1-\beta=0$ (Sect. \ref{sec:2_2_mergingtime}), or with a $L_2$ mass outflow with $\beta+\upsilon=1$ and $\epsilon=1-\beta-\upsilon=0$ (Sect. \ref{sec:L2_outflow}). We want to quantify the impact of including an efficiency $\epsilon >0$ in our calculations for the GW mergers boundaries. To this purpose, we are not considering here the additional source of loss from $L_2$ and therefore the efficiency for conservative MT is such that $\epsilon=1-\beta$. 

Both the common envelope and the time delay boundaries are expected to change when conservative MT plays a role. The Roche lobe mass-radius exponent is expected to vary smoothly between the limiting cases of $\beta=1$ (Eq. \ref{eq:zeta_iso}) and $\epsilon=1$ (Eq. \ref{eq:zeta_cons}), and is described in Eq. \ref{eq:zeta_uno_meno_beta}. In Fig. \ref{fig:zetas}, this translates into a smooth variation between the solid blue line (isotropic re-emission) and the solid golden line (fully conservative). As a consequence, a shift in critical mass ratios is also expected. In practice, we derived the modified critical mass ratios $q_{\mathrm{crit}}^{\epsilon=1-\beta}$ by solving
\begin{equation}\label{eq:stability_CONS_condition}
\zeta_{\mathrm{ad}}(R_{\mathrm{d,i}},m_{\mathrm{d,i}})=\zeta_{RL}^{\epsilon=1-\beta}(q_{\mathrm{crit}}^{\epsilon=1-\beta})\:.
\end{equation}
On the other hand, the period shrinkage (or widening) $P_{\mathrm{f}}/P_{\mathrm{i}}$ is also modified, varying smoothly between Eq. \ref{eq:ratio_isotropic_reemission} and the conservative limit (see, e.g. Eq. 33 in \citealt{Soberman1997}). By solving the equation of angular momentum balance, one can find
\begin{equation}\label{eq:ratio_CONS}
  \dfrac{P_{\mathrm{i}}}{P_{\mathrm{f}}}=\left(\dfrac{q_{\mathrm{f}}+1}{q_{\mathrm{i}}+1}\right)^{-\frac{3\beta\epsilon}{\epsilon(\epsilon-1)}-1} \left(\dfrac{\epsilon q_{\mathrm{f}}+1}{\epsilon q_{\mathrm{i}}+1}\right)^{\frac{3\beta}{\epsilon(\epsilon-1)}-5}\left(\dfrac{q_{\mathrm{f}}}{q_{\mathrm{i}}}\right)^3\:,
\end{equation}
where the final mass ratio $q=m_{\mathrm{d,f}}/m_{\mathrm{a,f}}$ takes into account that the final accretor mass is
\begin{equation}
m_{\mathrm{a,f}}=m_{\mathrm{a,i}}+\epsilon\:(1-\alpha_{\mathrm{core}})\:m_{\mathrm{d,i}}\:.
\end{equation}

\begin{figure}
   \resizebox{\hsize}{!}{
   \includegraphics[width=1.\textwidth]{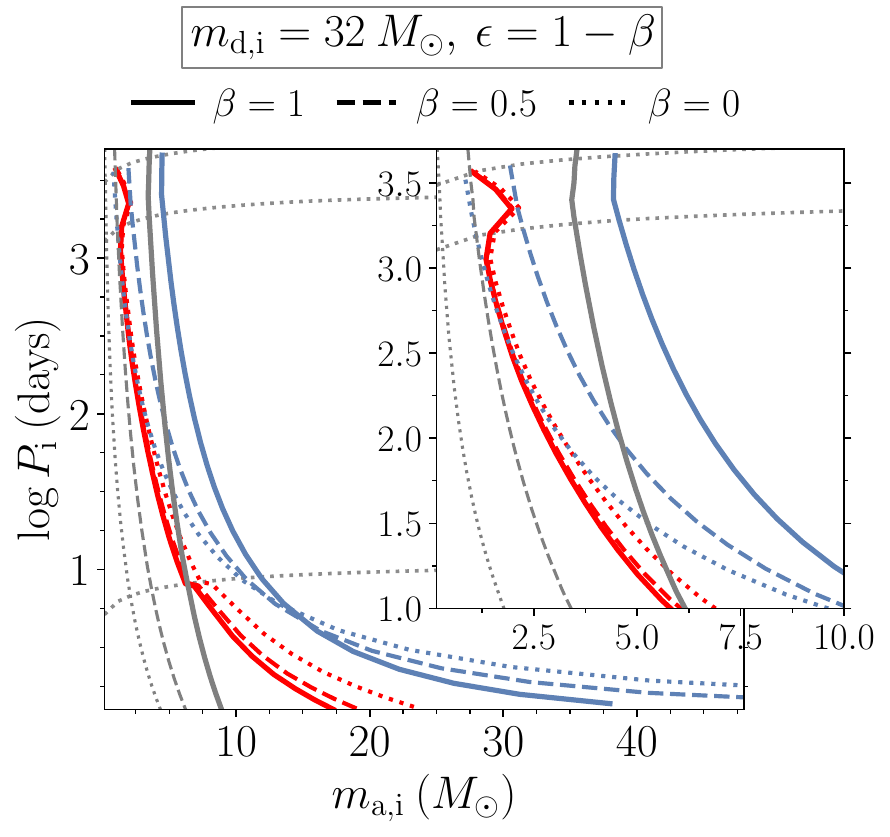}}
    \caption[]{{\label{fig:ConservativeL_32}} Fig. \ref{fig:mdonor32_L2}, but illustrating the impact of conservative MT. Boundaries are shown for the case of full isotropic re-emission with $\epsilon = 0$ (solid), $\epsilon = 0.5$ (dashed) and $\epsilon = 1$ (dotted). Same as Fig. \ref{fig:mdonor32_L2}. The efficiency for isotropic re-emission and conservativeness are varied such that $\epsilon=1-\beta$.}
   \end{figure}

Lastly, the boundary for $R_{\mathrm{d,f}}\leq R_{RL,\mathrm{f}}$ depends on the amount of shrinkage (widening) of the orbit, thus is also affected. In Fig. \ref{fig:ConservativeL_32} we show the results of our calculations for a donor star of $m_{\mathrm{d,i}}=32\:M_{\odot}$. We varied discretely the efficiencies for isotropic re-emission $\beta=0,0.5,1$ and conservative MT $\epsilon=1-\beta$. Including conservativeness leads to less shrinkage (and / or widening, if the mass ratio inversion is reached) of the orbit at the end of the MT episode; on the other hand, the accretor retains some transferred mass, so that the final binary is more massive. This explains the cut in parameter space at the longer initial periods $\log P_{\mathrm{i}}>1.5$ and the turnover of the delay time boundary (blue lines) at $\sim 15\:M_{\odot}$. Additionally, the Roche lobe boundary $R_{\mathrm{d,f}}<R_{RL,\mathrm{f}}$ becomes less stringent as the shrinkage is less extreme, and the instability boundary is mainly affected at the larger accretor masses, where the shift in $q_{\mathrm{crit}}^{\epsilon=1-\beta}$ is larger (see Fig. \ref{fig:zetas} - top panel).

Although we find that conservative mass transfer can still produce merging binary BHs, the situation is different for donor stars that produce NSs and WDs. The same analysis done for those lower mass donors indicate that the favorable strip for GW formation vanishes for all possible orbital periods. This implies that understanding the conservativeness of mass transfer with compact object accretors is critical.

\section{Comparison with observations}\label{sec:4_Comparison with observations}
\subsection{Single-degenerate binaries}\label{sec:3_4_comparison_known_systems}
Even if we expect a set of conditions for a single-degenerate system to lead to coalescing compact objects, it is not a given that those conditions are realized in nature. Whether the CO + star systems studied in Sect. \ref{sec:Results} are formed depends on the evolution of the initially more massive star and its potential interaction with its companion before the formation of the first CO of the system. To check for the existence of the necessary conditions in the single-degenerate stage we compare the properties of known systems at the CO + star stage to our theoretical predictions for GW progenitors' initial properties.

To this purpose we compile data of known CO + star systems from literature. For the High Mass (X-Ray) Binaries (HM(XR)Bs) with black holes we collect the data for the six known systems with characterised orbital properties and masses: HD 96670 (\citealt{Gomez2021}), M33 X-7 (\citealt{Pietsch2006}, \citealt{Orosz2007}), LMC X-1 (\citealt{Mark1969}, \citealt{Hutchings1983}, \citealt{Hutchings1987}), Cyg X-1 (\citealt{Miller-Jones2021}), VFTS 243 (\citealt{Shenar2022}), HD 130298 (\citealt{Mahy2022}), see Table \ref{tab:BHs} for a summary of these systems' properties. In this context, we call $m_{\mathrm{star}}$ the mass of the optical companion to the CO, $m_{\mathrm{CO}}$.

For HM(XR)Bs with neutron stars, we consider the galactic systems of Be/X-ray binaries where the CO is a NS, of which a sample of ~60-80 is known in the Milky Way (\citealt{Reig2011}). We collect the data listed in table 1 of \cite{Tomsick2010}, which summarizes the sample with estimates on the Be star masses $m_{\mathrm{star}}$.
To compute the ratio $m_{\mathrm{CO}}/m_{\mathrm{star}}$ of the neutron star mass versus the mass of the optical companion, we adopt the reference values for the masses shown in \cite{Tomsick2010}, without accounting for their uncertainties. 

For Low Mass X-ray Binaries (LM(XR)Bs), multiple examples exist, in literature, with white dwarfs objects as accretors. However, these are Roche filling / strongly interacting systems (e.g. cataclysmic variables, \citealt{Ritter2003} or classical novae, \citealt{Darnely2004}) where the current mass of the donor star does not represent its pre-interaction properties required in our analysis. Determinations of masses and periods for detached single-degenerate binaries with WDs are not commonly found in literature. We first consider systems composed by a WD + MS star with determined orbital properties: the Zwicky Transient Facility (ZTF) catalogue by \cite{Brown2023} for MS companions of spectral type M or later, i.e. ZTF systems, and the compilation by \cite{Holberg2013} for MS companions of spectral type K or earlier, i.e. the Sirius-like systems. 
We then considered systems composed by hot subdwarf stars of spectral type B or O (sdB/O) + MS companions (named ELCVNs from the prototype system, \citealt{Maxted2011}) and binaries with sdO + Be stars companions. These systems are assumed to be representative of a non-interacting stage preceding the single-degenerate configuration, as the subdwarf star is expected to evolve into a Helium WD. ELCVNs with determined mass ratios and orbital periods were taken from the catalogues in \cite{vanRoestel2018} (their table A4.2) and \cite{Maxted2014} (their table 2). To these, we added 10 systems found in the Kepler survey, with references in \cite{vanRoestel2018}. For sdO + Be/Xray stars, we used the compilation in \cite{Wang2023} (table 9). For each of these cases, we take the reference values for $q$ and $P_{\mathrm{orb}}$ without uncertainties.

\begin{figure}[h!]
      \resizebox{\hsize}{!}{
      \includegraphics[width=1.\textwidth]{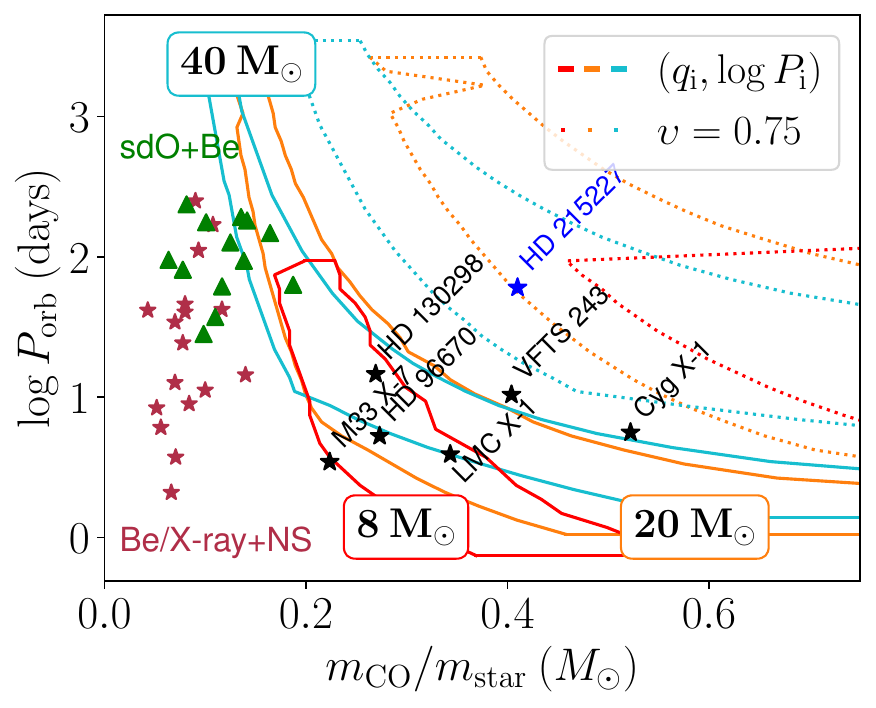}}
      \caption[]{{\label{fig:Xrays}} Observed orbital periods, i.e. $\log P_{\mathrm{orb}}$, and mass ratios, i.e. $m_{\mathrm{CO}}/m_{\mathrm{star}}$, for high (and intermediate) mass X-ray binaries and quiescent OB + BH systems. The markers report data points (uncertainties are neglected) listed in a) Table \ref{tab:BHs} for HM(XR)Bs with BHs (black), with the corresponding labeled names, b) table 1 from \cite{Tomsick2010} for HM(XR)Bs with NS (red). We also report the \cite{Casares2014} system (blue). The triangles are showing the sample of confirmed sdO + Be stars in table 9 from \cite{Wang2023}. Colored, solid contours are the correspondent loci predicted by our calculations to be promising for GW progenitors systems via stable MT, within the assumption of purely isotropic re-emission $\beta=1$. Every contour is labeled with the associated fixed donor's initial mass $m_{\mathrm{d,i}}$. The colored, dotted contours show the same loci $(m_{\mathrm{d,i}}/m_{\mathrm{a,i}},\log P_{\mathrm{i}})$ within the assumption of an efficiency for mass outflow via $L_2$ such that $\upsilon=0.5$.}
  \end{figure}

\begin{figure}[h!]
      \resizebox{\hsize}{!}{
      \includegraphics[width=1.\textwidth]{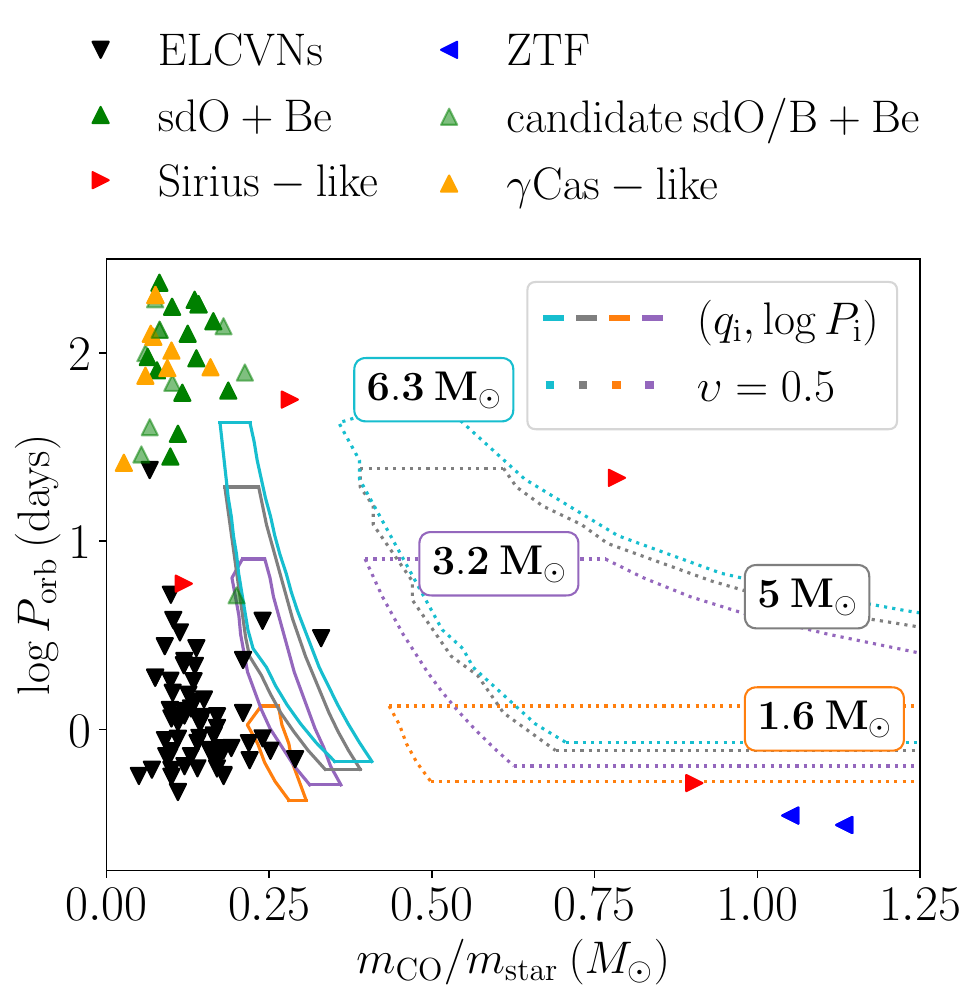}}
      \caption[]{{\label{fig:LOW_MASS}} Observed orbital periods, i.e. $\log P_{\mathrm{orb}}$, and mass ratios, i.e. $m_{\mathrm{CO}}/m_{\mathrm{star}}$, for low (and intermediate) mass single-degenerate binaries. More specifically: WD + MS star from the ZTF catalogue of \cite{Brown2023} (ZTF) and \cite{Holberg2013} (Sirius-like); sdO/B + MS star from \cite{vanRoestel2018}, \cite{Maxted2014} and 10 Kepler systems (see main text); sdO/B + Be stars are separated into candidate systems (candidate sdO/B + Be), confirmed sdO + Be and $\gamma$Cas-like systems as in \cite{Wang2023}. Markers report the reference data points without uncertainties. Colored, solid contours are the correspondent loci predicted by our calculations to be promising for GW progenitors systems via stable MT and follow the same notation as in Fig. \ref{fig:Xrays}. }
  \end{figure}

Figure \ref{fig:Xrays} shows the comparison between our theoretical predictions on the conditions needed to form a GW source (solid lines) and the six known HM(XR)Bs with BHs together with the considered sample of Be stars orbiting NSs and sdO + Be binaries (single markers). The population of HM(XR)Bs with BHs extends within $0\lesssim\log P_{\mathrm{orb}}(\mathrm{day})\lesssim 1$. The population of Be-Xray binaries with NSs lies within $0.05\leq m_{\mathrm{CO}}/m_{\mathrm{star}}\leq 0.18$, i.e. with mass ratios for which we predict mass transfer to be unstable. Similarly, the sample of sdO + Be stars lies in the unstable MT region (see also Fig. \ref{fig:LOW_MASS}), spanning larger periods $1.5\lesssim\log P_{\mathrm{orb}}(\mathrm{day})\lesssim 2.5$. The lack of lower mass Be stars in X-ray binaries (barring uncertainties in mass determinations from spectral types) has been used to argue for efficient MT in binary systems before the formation of the first CO (\citealt{Vinciguerra2020}). In this context, efficient MT is able to preclude the formation of Be/X-ray binaries with higher mass ratios than the ones that would evolve stably and form merging binary neutron stars. Conservative mass transfer prior to the first CO formation was also advanced as an explanation for the evolutionary stage of some sdO + Be star systems (e.g. HR6819, \citealt{Bodensteiner2020}). As for Be/X-ray systems with BHs, only one candidate is known, HD 215227 (MWC 656, \citealt{Casares2014}), also reported in Fig. \ref{fig:Xrays}. However, this system might contain a hot subdwarf object orbiting the Be-star (\citealt{Rivinius2022}).

Figure \ref{fig:Xrays} shows also the results switching on an efficiency for mass outflow from $L_2$ of $\upsilon =0.75$ (dashed lines). The predicted regions for GW progenitors parameter space are consistently shifted towards more extreme mass ratios, in an area which is not yet populated by known observations of CO + star binaries prior to the remaining star filling its Roche lobe. HD 215227 is the only system within the region for a massive donor predicted to directly collapse into a BH (the representative contour being the one for $m_{\mathrm{d,i}}=40\:M_{\odot}$), and at the same time close to the predicted region for a NS remnant (the representative contour being the one for $m_{\mathrm{d,i}}=20\:M_{\mathrm{\odot}}$).

Assuming that the post-MT state and the final fate of the optical companions for HM(XR)Bs are described by those of our theoretical results for the closest initial donor's mass found in the \cite{Ge2020} table, we can assess if the conditions for the stable mass transfer channel to produce a GW source are realized in nature. Some of these systems, however, present a non-negligible eccentricity, see Table \ref{tab:BHs}. Our predictions, on the other hand, are assuming that, at the stage of Roche lobe-filling donor, the CO + star system is already circular (we use Kepler's third law in the form of Eq. \ref{eq:common_envelope_boundary}).
To compare with our results for circular orbits we assume that, near the onset of Roche lobe overflow, the orbit circularizes to a period $P_{\mathrm{orb}}^{e=0}$ such that
\begin{equation}{\label{eq:equivalent_circular}}
P_{\mathrm{orb}}^{e=0}=P_{\mathrm{orb}}\times (1-e^2)^{2/3}\:,
\end{equation}
resulting from preservation of orbital angular momentum. For $e$ we used the values of the observed eccentricities in Table \ref{tab:BHs}.
We then associate each $m_{\mathrm{star}}$ to the closest initial donor's mass in our simulation set, $m_{\mathrm{d,i}}^{\mathrm{Ge}}$. We finally make a comparison, in Fig. \ref{fig:Xrays}, of the systems with the theoretical contours for different $m_{\mathrm{d,i}}^{\mathrm{Ge}}$.

Table \ref{tab:BHs_outcome} summarizes the predicted evolution (or outcome) of the known systems with BHs, basing on our purely isotropic re-emission, $\beta=1$, and mass outflow from $L_2$, $\upsilon=0.75$, evolutionary channels. We report the associated initial donor's mass in the \cite{Ge2020} data and the modeled fate of the second born CO. Lastly, we tabulate whether the systems have a set of initial properties that can lead to the evolution into merging double COs via stable MT (GW merger), or are expected to evolve into a wide binary (wide binary) or undergo unstable MT (unstable MT). 


Barring the limitations of this comparison, for the isotropic re-emission case $\beta=1$ we find systems covering all outcomes including unstable mass transfer, the formation of wide CO + CO systems and merging CO + CO expected to undergo stable MT (i.e., they appear within their associated theoretical contours in Fig. \ref{fig:Xrays}). This provides support that the conditions for the stable MT channel are realized in nature, provided that we neglect stellar winds widening the orbit after interaction (see Sect. \ref{sec:Effect of winds}). Interestingly, HD 96670 presents conditions that can be compatible with the formation of a NS-BH merger. Considering cases with $L_2$ mass outflow, $\upsilon >0$, the conditions to form a GW source shift to larger CO masses $m_{\mathrm{CO}}$. In particular, for $\upsilon=0.75$ we only find MWC 656 (for which the presence of a BH is contested, \citealt{Rivinius2022}) is compatible with the formation of a GW source. We also notice that two systems, i.e. VFTS 243 and M33 X-7, are not showing in Fig. \ref{fig:Xrays} the suitable properties for being GW progenitor systems via neither stable MT with $\beta=1$ nor with $\upsilon=0.75$. For VFTS 243, however, we can find a GW merger outcome via stable MT by switching on an efficiency for $L_2$ such that $0.25\lesssim\upsilon\lesssim 0.5$ (see Fig. \ref{fig:contours_L2_BHs}).

We show in Fig. \ref{fig:LOW_MASS} the same comparison as in Fig. \ref{fig:Xrays} between our theoretical predictions and the compilation of LM(XR)Bs and single-degenerate representative systems collected from literature (colored markers). The systems with sdO + Be stars are separated into candidate systems (candidate sdO/B + Be) from optical spectroscopy, confirmed sdO + Be and $\gamma$Cas-like systems (from the prototype \citealt{Secchi1866}), see table 9 in \cite{Wang2023} for the distinction. 
For the single-degenerate binaries WD + MS, consistent biases towards small orbital period ($\log P_{\mathrm{orb}}<0$) binaries (as well as low mass for the MS companion) are present in photometric surveys such as ZTF. On the other hand, Sirius-like systems are usually best resolved with proper motion surveys, biasing towards the wide orbits (mostly with $\log P_{\mathrm{orb}}>3$). The sample of ELCVNs and sdB/O + Be/Xray stars extends within $-0.1\lesssim\log P_{\mathrm{orb}}(\mathrm{day})\lesssim 2.5$ and mostly with mass ratios for which we predict mass transfer to be unstable. This holds true whether or not additional $L_2$ momentum loss is accounted for (dotted lines). The degenerate companions in ELCVNs systems are generally very low mass ($\sim 0.2\:M_{\odot}$) pre-Helium WDs, but more massive companions could shift the populated region to that predicted by us for GW mergers via stable MT.

\begin{table*}
\caption[mmm]{Summary of the six known HM(XR)Bs with black holes. 
}
\centering
{\renewcommand{\arraystretch}{1.1}
\begin{tabular}{lccccc}
\hline\hline 
Source    & $P_{\mathrm{orb}}$ (days)      & $m_{\mathrm{BH}}\:(M_{\odot})$ & $m_{\mathrm{star}}\:(M_{\odot})$ & $e$ & $q$  \\
\hline
\textbf{HD 96670}\tablefootmark{1}       & 5.28388$\pm$0.00046 & $6.2^{+0.9}_{-0.7}$            & $22.7^{+5.2}_{-3.6}$ &  $0.12\pm 0.01$           & 0.273         \\
\textbf{M33 X-7}\tablefootmark{2}      & 3.45301$\pm$0.00002 & $15.65\pm 1.45$                                & $70.0\pm 6.9$    & $0.0185\pm 0.0077$                             & 0.224       \\
\textbf{LMC X-1}\tablefootmark{3}          & $3.90917\pm 0.00005$               & $10.91\pm 1.41$                               & $31.79\pm3.48$    &  $0.0256\pm 0.0066$                            & 0.343          \\
\textbf{Cyg X-1}\tablefootmark{4}     & $5.599836\pm 0.000024$               & $21.2^{+2.2}_{-2.3}$                                & $40.6^{+7.7}_{-7.1}$ & $0.0189^{+0.0163}_{-0.0217}$                               & 0.522        \\
\textbf{VFTS 243}\tablefootmark{5}        & $10.4031\pm 0.0004$              & $10.1\pm 2.0$                               &  $25.0\pm 2.3$    & $0.017\pm 0.012$                           & 0.404          \\
\textbf{HD 130298}\tablefootmark{6}         & $14.62959\pm 0.000854$                & $>7.7\pm 1.5$             &  $28.0^{+5.2}_{-4.1}$         & 0.468                       & 0.275  \\  
\textbf{HD 215227}\tablefootmark{7}  & $60.37$ & $3.8 - 6.9$  & $10-16$ & $0.10\pm 0.04$  & $0.410$  \\
\hline
\end{tabular}
}
\label{tab:BHs}

\tablefoot{Columns 2 to 5 give the orbital periods $P_{\mathrm{orb}}$ in days, masses of the BH $m_{\mathrm{BH}}$, masses of the optical companion $m_{\mathrm{star}}$, and eccentricities $e$. The given mass ratio $q=m_{\mathrm{BH}}/m_{\mathrm{star}}$ is indicative and provided with no uncertainties (see main text).
\tablefoottext{1}{\cite{Gomez2021}.}
\tablefoottext{2}{\cite{Orosz2007}. }
\tablefoottext{3}{\cite{Orosz2009}. }
\tablefoottext{4}{\cite{Miller-Jones2021}. }
\tablefoottext{5}{\cite{Shenar2022}. For this system we do not have a measurement of the inclination, so there is a large range of possible masses for the BH; we report here the weighted mean, between their spectroscopic and evolutionary mass, i.e. $M_{1}$ of their Table 1.} 
\tablefoottext{6}{\cite{Mahy2022}. Similar to VFTS 243, we have a large range of possible $m_{\mathrm{BH}}$ for HD 130298; we thus use their lower limit on $m_{\mathrm{BH}}$ based on the binary mass function. We also use their evolutionary mass for the optical companion, see their appendix E.1.10. Their eccentricity value is also reported without uncertainty.} 
\tablefoottext{7}{\cite{Casares2014}. The orbital parameters (provided in their table 1) have been derived by \cite{Casares2014} by fixing the orbital period to the shown value here, which is thus reported without uncertainty.} 
}

\end{table*}

\begin{table*}
\caption{Predicted outcome of the second (2nd) born CO for the known HM(XR)Bs with black holes.} 
\centering
{\renewcommand{\arraystretch}{1.1}
\begin{tabular}{lccccc}
\hline\hline 

\multicolumn{1}{c}{Source} &
\multicolumn{1}{c}{$P_{\mathrm{orb}}^{e=0}$ (days)} &
\multicolumn{1}{c}{$m_{\mathrm{d,i}}^{\mathrm{Ge}}\:(M_{\odot})$}   &
\multicolumn{1}{c}{\text{2nd born CO}}              &
\multicolumn{2}{c}{\text{OUTCOME}}              \\ 
\hline
 & & &          &\multicolumn{1}{c}{$\beta=1$} & \multicolumn{1}{c}{$\upsilon=0.75$} \\ \cline{1-6}

\multirow{2}{*}{\text{HD 96670}}&\multirow{2}{*}{5.23} &20  & NS  & \multicolumn{1}{c}{GW merger} & \multicolumn{1}{c}{unstable MT}         \\
& &                   \multicolumn{1}{c}{25} & \multicolumn{1}{c}{BH}           & \multicolumn{1}{c}{GW merger }            & \multicolumn{1}{c}{unstable MT} \\
\hline
                   
\text{M33 X-7} & 3.4522 & 63    &  BH &\multicolumn{1}{c}{unstable MT} &\multicolumn{1}{c}{unstable MT}      \\
\hline
\text{LMC X-1} & 3.9075 & 32      & BH & \multicolumn{1}{c}{GW merger} &\multicolumn{1}{c}{unstable MT}       \\
\hline
\text{Cyg X-1} &5.5985 & 40      & BH & \multicolumn{1}{c}{wide binary}&\multicolumn{1}{c}{unstable MT}       \\
\hline
\text{VFTS 243} &10.401 & 25   & BH & \multicolumn{1}{c}{wide binary} &\multicolumn{1}{c}{unstable MT}       \\
\hline
\multirow{2}{*}{\text{HD 130298}} & \multirow{2}{*}{12.407} & 25   & BH  & \multicolumn{1}{c}{GW merger} & \multicolumn{1}{c}{unstable MT}\\
& &  \multicolumn{1}{c}{32} & BH & \multicolumn{1}{c}{GW merger} & \multicolumn{1}{c}{unstable MT}  \\
\hline
\multirow{2}{*}{\text{HD 215227}} & \multirow{2}{*}{59.97} & 10   & NS  & \multicolumn{1}{c}{wide binary} & \multicolumn{1}{c}{GW merger}\\
& &  \multicolumn{1}{c}{16} & NS & \multicolumn{1}{c}{wide binary} & \multicolumn{1}{c}{GW merger}  \\

\hline
\end{tabular}
}
\tablefoot{ The mass of the systems is assumed to be $m_{\mathrm{star}}=m_{\mathrm{d,i}}^{\mathrm{Ge}}$, where $m_{\mathrm{d,i}}^{\mathrm{Ge}}$ represents the initial donor's mass closest to the actual $m_{\mathrm{star}}$ in \cite{Ge2020} tables. Comparison to our results is made using the circularised orbital period $P_{\mathrm{orb}}^{e=0}$ as in Eq. \ref{eq:equivalent_circular}, for eccentric systems with $e>0$. The estimate of $P_{\mathrm{orb}}^{e=0}$ is reported with the same significant digits as the eccentricities in Table \ref{tab:BHs}. The hypothesized channels of purely isotropic re-emission, $\beta=1$, and mass outflow from $L_2$ with efficiency $\upsilon=0.75$ are both considered. The outcomes are also labeled to argue whether the systems have a set of initial properties $(m_{\mathrm{CO}}/m_{\mathrm{star}},P_{\mathrm{orb}}^{e=0})$ that can lead to the evolution into merging double COs via stable MT (GW merger), or are expected to evolve into a wide binary (wide binary) or undergo unstable MT (unstable MT).}
\label{tab:BHs_outcome}
\end{table*}


\subsection{Compact objects mergers}\label{sec:GWTC-3}
\begin{figure*}
   \resizebox{\hsize}{!}{
   \includegraphics[width=1.\textwidth]{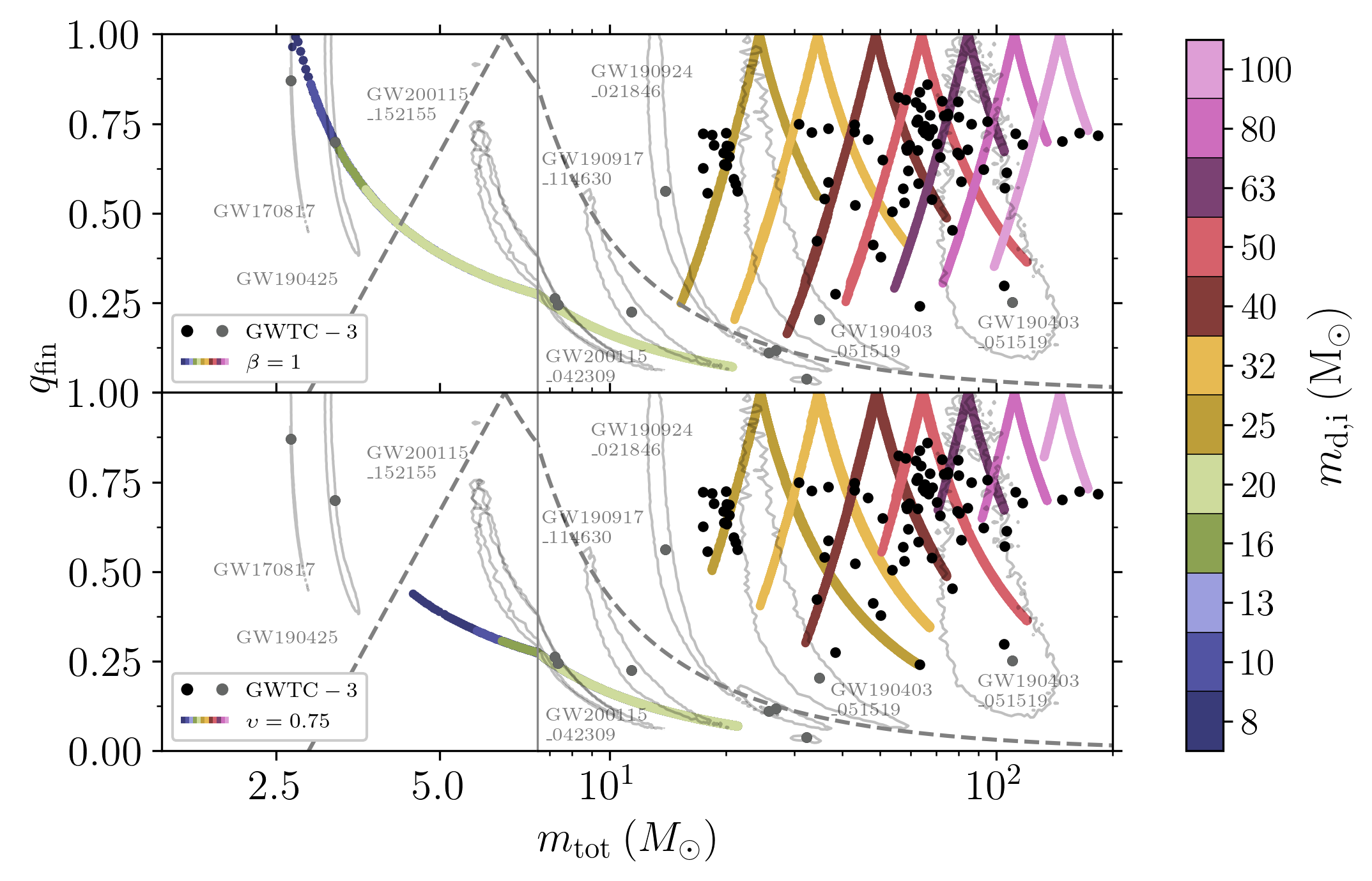}}
    \caption[]{Plane of final mass ratio $q_{\mathrm{fin}}<1$ and total final mass, $m_{\mathrm{tot}}$. We report the estimation values with corresponding 90\% credible intervals for the GWTC-3 signal (gray/black dots) and the posterior 90\% iso-contour for a selection of those (gray dots and light gray curves). Resulting final mass ratios and total mass from our treatment are reported with color coding referring to different donor masses $m_{\mathrm{d,i}}$ in the simulation grid from Ge et al. series \cite[\citeyear{Ge2015}, \citeyear{Ge2010}]{Ge2020}. The assumption of completely non-conservative, purely isotropic re-emission, $\beta=1$, channel is adopted in the top panel, while the bottom panel shows the results when switching on an efficiency for $L_2$ mass outflow of $\upsilon=0.75$. Solid gray, dashed lines delineate regions where the primary and secondary can have a mass below 3 $M_{\odot}$, and where both objects in the binary have masses above 3 $M_{\odot}$.}
    {\label{fig:GWs}}
   \end{figure*}

We want to compare the GWTC-3 information with the results, from our simulation set, about the mass ratios, $q_{\mathrm{f}}$, and total mass, $m_{\mathrm{f}}=m_{\mathrm{a,f}}+\alpha_{\mathrm{core}}m_{\mathrm{d,i}}$, achievable after the stable MT episode. To this purpose, Fig. \ref{fig:GWs} compares the total mass and mass ratios of detected compact object mergers against our predictions for stable MT. Together with the results from our study, we show the 90\% density interval as the credible regions for some representative systems to illustrate the uncertainties in total mass and mass ratio of the GW observations.

We notice a difference in the $(m_{\mathrm{tot}},q_{\mathrm{fin}})$ shapes between the donor masses in the range of BH progenitors, i.e. $m_{\mathrm{d,i}}\gtrsim 20\:M_{\odot}$, and the ones in the range of NS progenitors, i.e. $8\lesssim m_{\mathrm{d,i}}/M_{\odot}\lesssim 20$. As for the BH progenitors, for each given donor mass the shape extends till $q_{\mathrm{fin}}=1$, showing again that for all such donors the second born CO could be the more massive one (which happens at longer initial periods $P_{\mathrm{i}}$, see e.g. Fig. \ref{fig:contours_L2_BHs}), \textit{or} the less massive one (happening, conversely, at smaller initial periods $P_{\mathrm{i}}$ in Fig. \ref{fig:contours_L2_BHs}). For the NS progenitors, the shape does not reach the unitary final mass ratio (aside from a small portion of points in the case of $m_{\mathrm{d,i}}=8\:M_{\odot}$) because we assumed that the post-SN object would have a canonical mass of $1.4\:M_{\odot}$, see Sect. \ref{sec:NSdonor}, resulting in a final mass ratio $q_{\mathrm{f}}=1.4/m_{\mathrm{a,f}}$ mostly such that $q_{\mathrm{f}}<1$. Therefore, if the second born CO is a NS (i.e. coming from a SN explosion of a donor star such that $8\lesssim m_{\mathrm{d,i}}/M_{\odot}\lesssim 20$) of canonical mass, it will most likely be the less massive object of the CO + CO pair. 

We also observe that the stable MT channel, in the purely isotropic re-emission limit, has the potential to reproduce the final properties of the majority of GWTC-3 signals, in terms of final mass ratios and total mass of the mergers. Covered mass ratios are such that $q_{\mathrm{fin}}\gtrsim 0.1$. On the other hand, our predictions cover the full range of final total masses $m_{\mathrm{tot}}$ (although we do not include the effect of pair instability SNe in our treatment of BH progenitors, which could limit the possible ranges). Notable exceptions are GW191219-163120, which stands at $q_{\mathrm{fin}}= 0.04^{+0.01}_{-0.01}$ and $m_{\mathrm{tot}}= 32.3^{+2.2}_{-2.7}\:M_{\odot}$; GW200210-092254 at $q_{\mathrm{fin}}= 0.12^{+0.05}_{-0.04}$ and $m_{\mathrm{tot}}= 27.0^{+7.1}_{-4.3}\:M_{\odot}$; GW190814 at $q_{\mathrm{fin}}= 0.11^{+0.03}_{-0.02}$ and $m_{\mathrm{tot}}=25.9^{+1.3}_{-1.3}\:M_{\odot}$. 

Including higher angular momentum loss by switching on an efficiency $\upsilon >0$ for $L_2$ outflow (Fig. \ref{fig:GWs} - bottom) results in less extreme final mass ratios ($q_{\mathrm{fin}}\gtrsim 0.25$ for $\upsilon=0.75$, as opposed to $q_{\mathrm{fin}}\gtrsim 0.1$ for $\beta=1$) coming from merging binary black holes, and no possible path to the formation of a merging binary neutron star from stable mass transfer (see section Sect. \ref{sec:NSdonor}). This is due to the common envelope boundary being shifted to smaller final mass ratios $q_{\mathrm{f}}$ at the end of the MT episode, making any mass transfer event from a hydrogen rich NS progenitor onto a NS unstable (see e.g. Fig. \ref{fig:mdonor8_L2}).

An important limitation on this comparison with the population of observed GWTC-3 CO + CO mergers comes from the fact that we are not considering the information on spins. The high inferred BH spins in HM(XR)Bs have been used to argue that they are subdominant progenitors for the observed population of binary BH mergers (e.g. \citealt{Fishback2022}, \citealt{GallegosGarcia2022}), while others consider spin measurements in HM(XB)s should be taken with care (\citealt{Belczynski2021}). Among the models that attempt to explain the near critical spins inferred for HM(XR)Bs, one idea is that the spin is inherited from their progenitor star being tidally locked in a short period orbit during its main sequence (\citealt{Valsecchi2010}, \citealt{Qin2019}). If this is the case, longer period OB + BH systems would be expected to have lower black hole spins potentially consistent with the observed binary BH population. The detections of VFTS 243 (\citealt{Shenar2022}) and HD 130298 (\citealt{Mahy2022}) have started to probe these longer orbital periods and the Gaia mission is expected to provide hundreds of additional detections in this regime (e.g. \citealt{Breivik2017}, \citealt{Mashian2017}, \citealt{Wiktorowicz2019}, \citealt{Janssens2022}, \citealt{Janssens2023}). However, such wide systems are expected to lack accretion disks (\citealt{Sen2021}, \citealt{Hirai2021}), making it impossible to apply current methods to measure BH spins.


\section{Conclusions}\label{sec:Conclusions}

We studied the parameter space of initial properties of systems composed by a first born CO orbiting a Roche lobe-filling donor. At an arbitrary evolutionary stage, the system is expected to undergo a stable MT episode and, after the formation of the second born CO, to evolve into a double compact object configuration able to merge within a Hubble time. We summarize our main findings in the following: 

    \begin{enumerate}
    \item Under the assumption of fully non-conservative mass transfer and isotropic re-emission from the accreting compact object, we find that stable mass transfer can produce all possible pairings of merging double compact objects except for WD + BH binaries. This holds barring the limitations on the properties (masses and orbital periods) of single-degenerate binaries that arise from the evolution until the formation of the first compact object.
    
    \item The possible formation of NS + NS mergers from stable MT is limited to a narrow range in initial properties. For a typical neutron star mass of 1.4 $M_{\odot}$, any companion massive enough to form a second born NS ($m_{\mathrm{d,i}}\gtrsim 8\:M_{\odot}$) will have an extreme mass ratio, such that binary NSs can only be produced in a narrow range of donor masses and orbital periods.
    
    \item Including additional angular momentum loss (by considering a fraction of the mass removing the angular momentum corresponding to $L_2$), the initial orbital periods suitable for GW mergers via stable MT are shifted towards larger initial mass ratios $q=m_{\mathrm{d,i}}/m_{\mathrm{a,i}}$. Such additional angular momentum loss can prevent stable MT to form merging binary NSs. On the other hand, the inclusion of $L_2$ mass outflow with efficiency $\geq$ 50\% can allow a WD + BH pair to merge within the age of the universe.
    
    \item We compared our results with observations of single-degenerate binaries, i.e. HM(XR)Bs with black holes and Be-Xray binaries with neutron stars. For the isotropic re-emission calculations, the required conditions for the stable MT channel to operate seem to be present in nature. In particular, the candidate BH system HD 96670 (\citealt{Gomez2021}) has properties consistent with the formation of a merging NS + BH. As future (and current) surveys are expected to unveil more high mass BHs, in particular the wider orbits counterparts for BH companions and the lower mass optical companions of NS, we can expect to observe more systems showing such favorable conditions.
    
    \item Considering isotropic re-emission, stable mass transfer can produce merging binary black holes with mass ratios $q\geq 0.1$. This covers the majority of the sources observed in the GWTC-3 catalogue, except for extreme cases such as GW190814. Considering enhanced angular momentum loss limits the minimum mass ratios that can be achieved. In particular, under the assumption that 75\% of the material is removed with the angular momentum of the $L_2$ point, we do not find BH mergers with $q<0.25$.
    \end{enumerate}
    
We make our results available to the community in tabular, machine readable form on Zenodo (\href{doi.org/10.5281/zenodo.8334056}{doi.org/10.5281/zenodo.8334056}): tables at fixed donor mass $m_{\mathrm{d,i}}$ listing the boundaries $(m_{\mathrm{a,i}},\log P_{\mathrm{i}})$ for GW progenitor systems at different efficiencies of angular momentum loss, i.e. varying discretely $\beta\in [0,1]$ and $\upsilon\in [0,1]$ and keeping the total efficiency $\epsilon=\beta+\upsilon=1$. Modified boundaries with some degree of conservativeness are also provided, in the case $\epsilon=1-\beta$, for $\beta=0, 0.5, 1.$.

Our work provides a simple semi-analytical approach to study the physics and limitations of the stable mass transfer channel. We therefore implicitly assume the existence of specific configurations of CO + star systems. Although we have shown that comparison to known systems in that intermediate stage is promising, it is important to validate our results using detailed binary evolution calculations, which we aim to do in future work. Simulations starting from two stars at the ZAMS, rather than after the formation of the first CO, are also necessary to determine the formation rate of these systems (e.g. \citealt{Bavera2021}, \citealt{GallegosGarcia2021}) and compare to observed rates of CO coalescences. Detailed simulations can also show the impact of a previous mass transfer phase onto the donor star in a CO + star system, which will modify its response to mass transfer and thus its stability to it.

\begin{acknowledgements}
     AP acknowledges support from the FWO PhD fellowship fundamental research under project 11M8323N and the Interuniversity BOF project No.20-VLIR-iBOF-021. PM acknowledges support from the FWO junior postdoctoral fellowship No. 12ZY520N and the FWO senior postdoctoral fellowship No. 12ZY523N. This project has received funding from the European Research Council (ERC) under the European Union’s Horizon 2020 research and innovation programme (grant agreement No. 772225/MULTIPLES). GN acknowledges support form the Dutch Science Foundation NWO.
\end{acknowledgements}

%

%
%

\bibliographystyle{aa} 
\bibliography{Forming_merging_double_compact_objects_with_stable_mass_transfer} 

\begin{appendix} 

\section{$L_2$ calculations}\label{sec:appB}
To determine the change in orbital period due to an outflow from the $L_2$ point one needs to compute the integral (see Section \ref{sec:L2_outflow})
\begin{eqnarray}
\int_{q_\mathrm{i}}^{q_\mathrm{f}}\frac{1+q}{q}\left[x_{L_2}(q)-x_{\mathrm{cm}}\right]^2 dq=\int_{q_\mathrm{i}}^{q_\mathrm{f}}f(q)\: dq,
\end{eqnarray}
where $f(q)$ is defined as
\begin{eqnarray}
    f(q)=&\displaystyle \frac{1+q}{q}\:\left[1+\sum_{n=1}^5 a_n\times
    \begin{cases}
        q^{-(n-n_0)} & q\geq1 \\[5pt]
        q^{(n-n_0)} &  q<1
    \end{cases}
    \right]\\
    \equiv &
    \begin{cases}
        f_>(q) & q\geq1 \\[5pt]
        f_<(q) & q<1
    \end{cases}\qquad\qquad\qquad\quad
\end{eqnarray}
\[\mathrm{with}\hspace{0.25cm}n_0=0.658,\]
\[a_1=1.544,\:a_2=-3.118,\:a_3=4.387,\:a_4=-3.567,\:a_5=1.190\:.\]
In the above we made use of the fit to $x_{L_2}-x_{\mathrm{cm}}$ in Eq. \ref{eq:L2_fit}. We also introduced the functions $f_{\gtrless}(q)$ to describe the piece-wise function $f(q)$ in the mass ratio regimes $q\gtrless 1$, respectively. As the integrand $f(q)$ is a piecewise function, we need to consider different possible outcomes depending on whether or not the binary crosses a mass ratio of unity during the MT event. Let us first consider the indefinite integral $I(q)$
\begin{equation}
I(q)=
\begin{cases}
    I_{>}(q)\equiv\int\:f_>(q)\:d q & q\geq 1 \\[5pt]
    I_{<}(q)\equiv\int\:f_<(q)\:d q & q< 1
\end{cases}.
\end{equation}
By direct integration we have that
\begin{equation}\label{eq:L2_primitive_definitive}
I(q)=\begin{cases}
    I_{>}(q)=q+\log q +\mathlarger{\sum}_{n=0}^{n_{\mathrm{max}}} a^{>}_n\: q^{-(n-n_0)} & q\geq 1 \\[15pt]
    I_{<}(q)=q+\log q +\mathlarger{\sum}_{n=0}^{n_{\mathrm{max}}} a^{<}_n\: q^{(n-n_0+1)} & q< 1
\end{cases}
\end{equation}
with
\[a^>_n=-\dfrac{a_n+a_{n+1}}{n-n_0}\:,\hspace{0.5cm}a^<_n=\dfrac{a_n+a_{n+1}}{n-n_0+1}\:,\]
and the coefficients $a_n$ and $n_0$ are given in Eq. \ref{eq:L2_fit}. In this expression $n_{\mathrm{max}}=5$ and $a_0=0=a_{n>5}$. The result of the definite integral can then be computed as
\begin{equation}\label{eq:L2_integral_pieces}
\int_{q_{\mathrm{i}}}^{q_{\mathrm{f}}}f(q)\:d q=
\begin{cases}
I_{>}(q_{\mathrm{f}})-I_>(q_{\mathrm{i}}) & q_{\mathrm{f}}>1,q_{\mathrm{i}}\geq 1 \\[5pt]
I_<(q_{\mathrm{f}})-I_<(1)+I_>(1)-I_>(q_{\mathrm{i}}) & q_{\mathrm{f}}<1,q_{\mathrm{i}}\geq 1 \\[5pt]
I_<(q_{\mathrm{f}})-I_<(q_{\mathrm{i}}) & q_{\mathrm{f}}<1,q_{\mathrm{i}}<1
\end{cases}.
\end{equation}

\section{Comparison to Marchant et al. (2021) results}\label{app:Marchant_comparison}
\begin{figure}
   \centering
   \includegraphics[width=0.4725\textwidth]{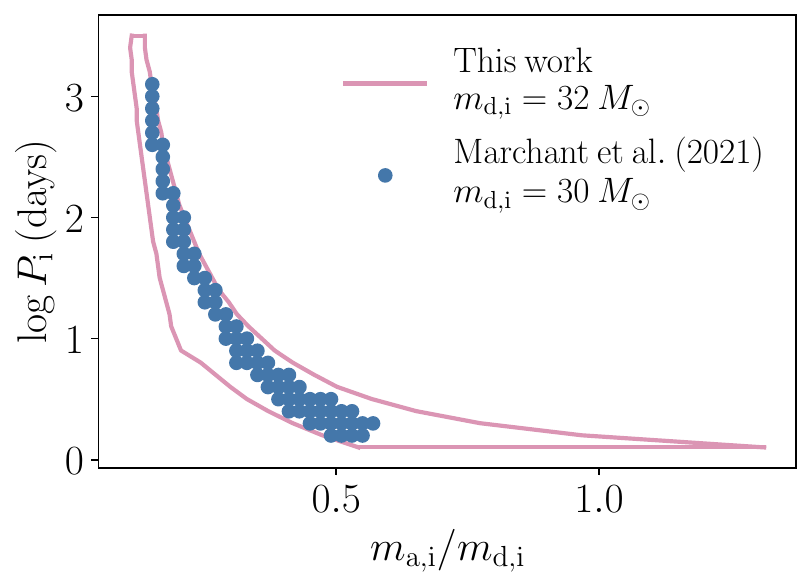}
      \caption[]{Parameter space ($q_0=m_{\mathrm{a,i}}/m_{\mathrm{d,i}}-\log P_{\mathrm{i}}$) of GW progenitors initial properties, comparing results from the simulations from \cite{Marchant2021} (blue dots), for a donor of $m_{\mathrm{d,i}}=30\:M_{\odot}$ and results from this work (pink contour), for a donor of $m_{\mathrm{d,i}}=32\:M_{\odot}$. }
      {\label{fig:PabloCOMP}}
   \end{figure}

As this work serves as a benchmark and is meant to be followed by a detailed set of \texttt{MESA} simulations, we also want to compare our results for a donor of $m_{\mathrm{d,i}}=32\:M_{\odot}$ with the ones presently available. To this purpose, we consider the work from \cite{Marchant2021}, in which the evolution of a metal poor ($Z=0.00142$) 30 $M_{\odot}$ donor star orbiting a CO is modeled with \texttt{MESA}. For the comparison, we show in Fig. \ref{fig:PabloCOMP} the $(m_{\mathrm{a,i}}/m_{\mathrm{d,i}}-\log P_{\mathrm{i}})$ parameter space. Their grid of simulations spans initial orbital periods $P_{\mathrm{i}}$ ranging from $-0.3<\log P_{\mathrm{i}}(\mathrm{days})<3.5$, and initial mass ratios $m_{\mathrm{a,i}}/m_{\mathrm{d,i}}$ between 0.01 and 0.59 (see their paper, Sect. 3.1 for more details on the grid), translating into CO masses $0.3<m_{\mathrm{a,i}}/M_{\odot}<17.7$. 

The delay time boundary computed by us in the isotropic re-emission channel is close to the one found by Marchant et al., in which the degree of non-conservativeness is not a free parameter (i.e., they are not fixing an efficiency $\beta$). 
The efficiency of mass transfer is instead determined by limiting the accretion rate into the CO by its Eddington limit, with all the surplus material being removed under the assumption of isotropic re-emission. For the 30 $M_{\odot}$ donor considered by \cite{Marchant2021}, mass transfer rates exceed the Eddington limit of the compact object by orders of magnitude, resulting on a near zero efficiency. The simulations of \cite{Marchant2021} also include a treatment for outflows from the outer Lagrangian points (under situations where the donor grows significantly beyond its Roche lobe); these outflows account for only a small fraction of the total ejected mass (see their figure 13). Considering this, their simulations can be compared to our results with $\beta=1$.

The instability boundary in our study is pushed to lower mass ratios compared to the results of detailed simulations in \cite{Marchant2021}. Our instability boundary for $m_{\mathrm{d,i}}=32\:M_{\odot}$ is much influenced by the hard limit on $R_{\mathrm{He-ZAMS}}<R_{RL,\mathrm{f}}$, see Sect. \ref{sec:Final_orbital_separation}, and a realistic condition can be even more conservative. This apparent discrepancy might therefore attenuate or disappear, once one resolves the boundaries with detailed simulations.


   \end{appendix}

\end{document}